\documentclass[10pt, a4paper, oneside]{article}
\usepackage[hidelinks]{hyperref}
\usepackage{jucs2e}
\usepackage{graphicx}
\usepackage{url}
\usepackage{ulem}
\usepackage{mathtools}
\usepackage{scalerel}
\usepackage{setspace}
\usepackage[strict]{changepage}
\usepackage[compatibility=false]{caption}
\usepackage[protrusion=true, nopatch=footnote]{microtype}
\usepackage{fontspec}
\usepackage{afterpage}
\usepackage{ragged2e}
\usepackage{tabularx}
\usepackage{subcaption}
\usepackage{longtable}
\usepackage{pdflscape}   
\usepackage{natbib}
\usepackage{array}
\usepackage{booktabs}
\usepackage{multirow}
\usepackage{float}
\usepackage{longtable}
\usepackage{rotating}
\usepackage{array}

\usepackage{ragged2e}
\usepackage{tabularx}
\usepackage{pdfpages}
\usepackage{tabularray}
\usepackage{multirow}
\usepackage[pass]{geometry}

\usepackage{pdflscape}   
\usepackage[table]{xcolor}

\usepackage{fontspec}
\newcolumntype{L}[1]{>{\raggedright\arraybackslash}p{#1}}
\newcolumntype{C}[1]{>{\centering\arraybackslash}p{#1}}

\usepackage{titlesec}

\titleformat*{\section}{\Large\bfseries}
\titleformat*{\subsection}{\normalsize\bfseries}

\graphicspath{{./figures/}}

\usepackage[format=plain, labelfont=it, textfont=it, justification=centering]{caption}
\usepackage{breakcites}

\apptocmd{\frame}{}{\justifying}{}

\newcommand\jucs{{Journal of Universal Computer Science}}
\newcommand\jucsvol{vol. 33, no. 6 (2027)}
\newcommand\jucspages{2987-2989}
\newcommand\jucssubmitted{26/05/2026}
\newcommand\jucsaccepted{1/10/2026}
\newcommand\jucsappeared{n.d}
\newcommand\jucslicence{ CC BY 4.0}
\newcommand\startingPage{2987}
\newcommand\paperauthor{{...}}

\newcommand\papertitle{Responsible Institutional Analytics: Interpreting Bias with AI Support}

\header{\paperauthor \papertitle}

\begin{document}

\title{{\fontsize{14pt}{14pt}\selectfont{\vspace*{-3mm}\papertitle\vspace*{-1mm}}}}

\author{{\bfseries\fontsize{10pt}{10pt}\selectfont{Francielle Marques}} \\
   {\fontsize{9pt}{12pt}\selectfont{(Pompeu Fabra University, Barcelona, Catalonia\\
   \orcid{0000-0002-1992-8728}, 
   francielle.marques@upf.edu)}}
   \and
   {\bfseries\fontsize{10pt}{10pt}\selectfont{Ariel Ortiz-Beltrán}}\\
   {\fontsize{9pt}{12pt}\selectfont{(Pompeu Fabra University, Barcelona, Catalonia \\
   \orcid{0000-0003-1522-2362}, 
   ariel.ortiz@upf.edu)}}
   \and
   {\bfseries\fontsize{10pt}{10pt}\selectfont{Ishari Amarasinghe}}\\
   {\fontsize{9pt}{12pt}\selectfont{(Pompeu Fabra University, Barcelona, Catalonia \\
   \orcid{0000-0003-2960-4804}, 
   ishari.amarasinghe@upf.edu)}}
    \and
   {\bfseries\fontsize{10pt}{10pt}\selectfont{Davinia Hernández-Leo}}\\
   {\fontsize{9pt}{12pt}\selectfont{(Pompeu Fabra University, Barcelona, Catalonia \\
   \orcid{0000-0003-0548-7455}, 
   davinia.hernandez-leo@upf.edu)}}
}

\label{first}
\maketitle

{\fontfamily{ptm}\selectfont
\begin{abstract}
{\fontsize{9pt}{9pt}\selectfont{\vspace*{-2mm}
Institutional Analytics (IA) dashboards inform decision-making in higher education, yet data limitations, constraints in analytical techniques, and missing contextual information often affect their interpretation. To support more responsible interpretation of IA, we introduce FACTRIA, a framework that organizes potential biasing factors across four areas: the analytics pipeline, institutional context, course-level characteristics, and demographics. We used the FACTRIA framework as input to a generative-AI chatbot designed to prompt users to reflect on these factors while analyzing IA. A qualitative study with stakeholders, drawing on four authentic IA cases, and a transition network analysis showed that the chatbot prompted participants to recognize how overlooked factors influenced their initial interpretation. Findings indicated that combining a structured framework with AI-based guidance can enhance context-aware, responsible interpretation of institutional data.
}}
\end{abstract}}

{\fontfamily{ptm}\selectfont
\begin{keywords}
{\fontsize{9pt}{9pt}\selectfont{
Institutional Analytics, Generative AI, Transition Network Analysis, Dashboards}}
\end{keywords}}

{\fontfamily{ptm}\selectfont
\begin{category}
{\fontsize{9pt}{9pt}\selectfont{
I.2, L.3, L.3.1}}
\end{category}}

{\fontfamily{ptm}\selectfont
\begin{doi}
{\fontsize{9pt}{9pt}\selectfont{
10.3897/jucs.200690}}
\end{doi}}

\section{Introduction}
\label{introduction}

Institutional Analytics (IA) is a field that helps higher education institutions report data and insights to inform and improve decision-making related to teaching and learning \citep{yanosky2015analyticslandscape}. Drawing on indicators from sources such as learning management systems, student performance, and student satisfaction surveys, it enables stakeholders (academic managers, educators) to make evidence-based decisions to enhance educational quality and student outcomes \citep{paz2019academic}. IA includes Learning Analytics (LA)—which examines learners’ interactions, engagement, and progression within courses—while extending these insights across programs, departments, and cohorts \citep{Jones2019LAHE} to inform institutional systems, strategic planning, and stakeholders' actions.

As IA matures in higher education, research has shown that responsible interpretation depends not only on the indicators provided in IA dashboards but on users’ awareness of the potential biases embedded in the analytical context and processes \citep{Williamson2022Review}. Biases can emerge from analytical choices, characteristics of the underlying data, the data context, and how analysts frame or communicate results; each of these factors can influence how stakeholders interpret patterns and act on them \citep{socialdatabias}. Responsible interpretation, therefore, requires careful attention to sense-making, transparency about assumptions and limitations, and explicit consideration of factors that clarify findings, support comparability, and keep interpretations appropriately bounded by the data. Recent educational AI research further suggests that conversational scaffolding and reflective prompting can support users’ metacognitive engagement when interacting with complex information environments, although such systems require careful design to avoid reinforcing automation bias or uncritical acceptance of generated content \citep{shi2026llmreview}

Indeed, some frameworks have already been proposed for the adoption of LA in institutions that call for responsible considerations. For example, \citep{HELACMM} shows the importance of considering factors such as teaching, learning, and assessment, as well as data management and analytics, ethics, privacy, legal issues, and people competencies in the implementation of IA. Another framework, proposed by \cite{DELICATEframework}, developed a checklist, named DELICATE, with eight action points that institutional managers and decision-makers should consider to facilitate the trusted implementation of IA. Furthermore, \cite{hernandez2019analyticslearningdesign} proposed Analytics Layers for Learning Design, which identifies different data sources relevant to informed, data-driven decision-making in educational design. However, these frameworks remain generic and do not identify or address the specific sources of potential bias in the data context and analysis that may affect their responsible interpretation.

This paper addresses this gap by examining how to support stakeholders in interpreting IA responsibly, with particular attention to biasing factors that arise from the data context and analytical process. It proposes extending IA dashboards with mechanisms that surface existing knowledge about these factors and with a scaffolding layer that guides users towards a factor-aware, responsible interpretation. 

Specifically, the paper introduces a framework that integrates the currently fragmented knowledge on biasing factors in IA-FACTRIA (Framework of Factors in the Analytical process and the data Context for Responsible Institutional Analytics)—and presents a conversational-agent–based system that leverages GenAI with FACTRIA-aware retrieval-augmented generation.
Rather than offering interpretations or recommending decisions, the FACTRIA-aware chatbot prompts users to reflect on data quality, assumptions, comparability, and other potential biasing factors that may shape the interpretation of IA.
We illustrate the framework's applicability with four real IA cases at a higher education institution and assess the system's use with key stakeholders using transition network analysis \citep{TNAframework}. 

Throughout this paper, we use responsible interpretation of IA to refer to the process of drawing conclusions from visualizations while accounting for the factors that condition what those visualizations can legitimately support. We operationalize it through the FACTRIA framework (Section \ref{factria}), whose dimensions specify the factors to consider when interpreting a visualization.

The remainder of this paper is organized as follows. Section \ref{background} overviews the background of this work. Section \ref{factria} the FACTRIA Framework. Section \ref{sec:methods} presents the methods. Section \ref{results} presents the results.  Section \ref{discussionsConclusions} discusses the results, Section \ref{limitations} discusses the limitations, and Section \ref{conclusions} presents our conclusions.

\section{Background}
\label{background}

IA has become not only highly relevant but also increasingly useful in higher education, supporting evidence-informed decision-making around teaching quality, student experience, and institutional effectiveness \citep{analyticsHEI}. However, prior studies show that effective use of IA depends on understanding the sensitivity of its indicators arising from both the analytical process and the data context \citep{radchenko2020biasesmultifaceted, whatarewemeasuring}. For example, educational research identifies student satisfaction metrics as key IA indicators for evaluating teaching quality, yet also cautions that these metrics must be interpreted within context, as they may be influenced by demographic or contextual factors—such as gender \citep{professorAgeandGender, measuringgenderbias} or institutional policy conditions (e.g., those associated with COVID-19) \citep{AreTheyRelatedtoWhatStudentsLearn, beltran2022surviving}.

Aspects of the learning design are also relevant when interpreting indicators of student outcomes \citep{RIENTIES2016333across151modules, li2016modelling}. In particular, visual dashboards can help teachers gain insight into their students' knowledge, making these outcomes more interpretable and actionable \citep{calvo2019improving}. Additionally, research highlights that satisfaction with teaching materials, assessments, and workload correlates positively with the overall teaching experience \citep{li2016modelling, beyondbias}. In IA studies focusing on retention, predictive models have shown that dropout risks vary by discipline—being higher in Engineering and Architecture—and are influenced by personal habits, institutional experiences, and educational goals, necessitating stakeholder-inclusive agendas \citep{educationaldatamining, IAagenda}.

Moreover, research indicates that LA dashboards often exceed users’ interpretive capabilities, in part because they do not explicitly articulate the analytical-process factors that users must consider for responsible interpretation \citep{Lem2014, Pozdniakov2025, Susnjak2022}. When factors remain implicit,  users may overlook limitations, misjudge comparability, or draw conclusions that are insufficiently grounded in the underlying evidence. 

Thus, effective IA tools require not only clear visualizations but also support that helps users attend to the analytical and contextual conditions required for responsible interpretation. 
The literature similarly emphasizes the value of evaluation-focused approaches that help interpretation lead to responsible action (e.g., the Analytics4Action Evaluation Framework, validated at scale) while keeping claims proportionate to evidence \citep{Rienties2016, Rienties2018}. Consistent with accounts that expose modeling choices rather than treating algorithms as substitutes for judgment \citep{AtkissonWiley2011, modernFramework}. 

Regarding AI-powered tools for improving visualization readability, \cite{VizChat} developed VizChat, a prototype chatbot that generates AI-driven explanations for data visualizations. The system combines multimodal generative AI with retrieval-augmented generation to produce contextually relevant explanations and insights, as shown in a case study. Similarly, \cite{ChatLAD} employed multi-agent retrieval-augmented generation to generate explanations for learning analytics dashboards, tailoring them to user needs and enhancing teachers’ comprehension of classroom discourse analysis. In developing the FACTRIA framework, we drew not only on AI features but also on knowledge of the factors that hinder responsible interpretation of IA visualizations.

Additionally, existing frameworks for responsible analytics address related concerns from different perspectives. Governance frameworks such as DELICATE \citep{DELICATEframework} and SHEILA \citep{SHEILA} operate primarily at the institutional level, specifying principles related to issues such as consent and transparency. Approaches to visualization literacy, such as CALVI \citep{CALVI}, focus on graphical features that may mislead readers. In contrast, the FACTRIA framework specifies the factors that readers should consider when drawing conclusions from an institutional analytics (IA) visualization. Unlike these previous approaches, FACTRIA is operationalized at the moment of interpretation, providing the basis for our FACTRIA-aware chatbot.

\section{FACTRIA framework}
\label{factria}

\begin{figure}[!ht]
    \centering
    \includegraphics[width=0.50\columnwidth]{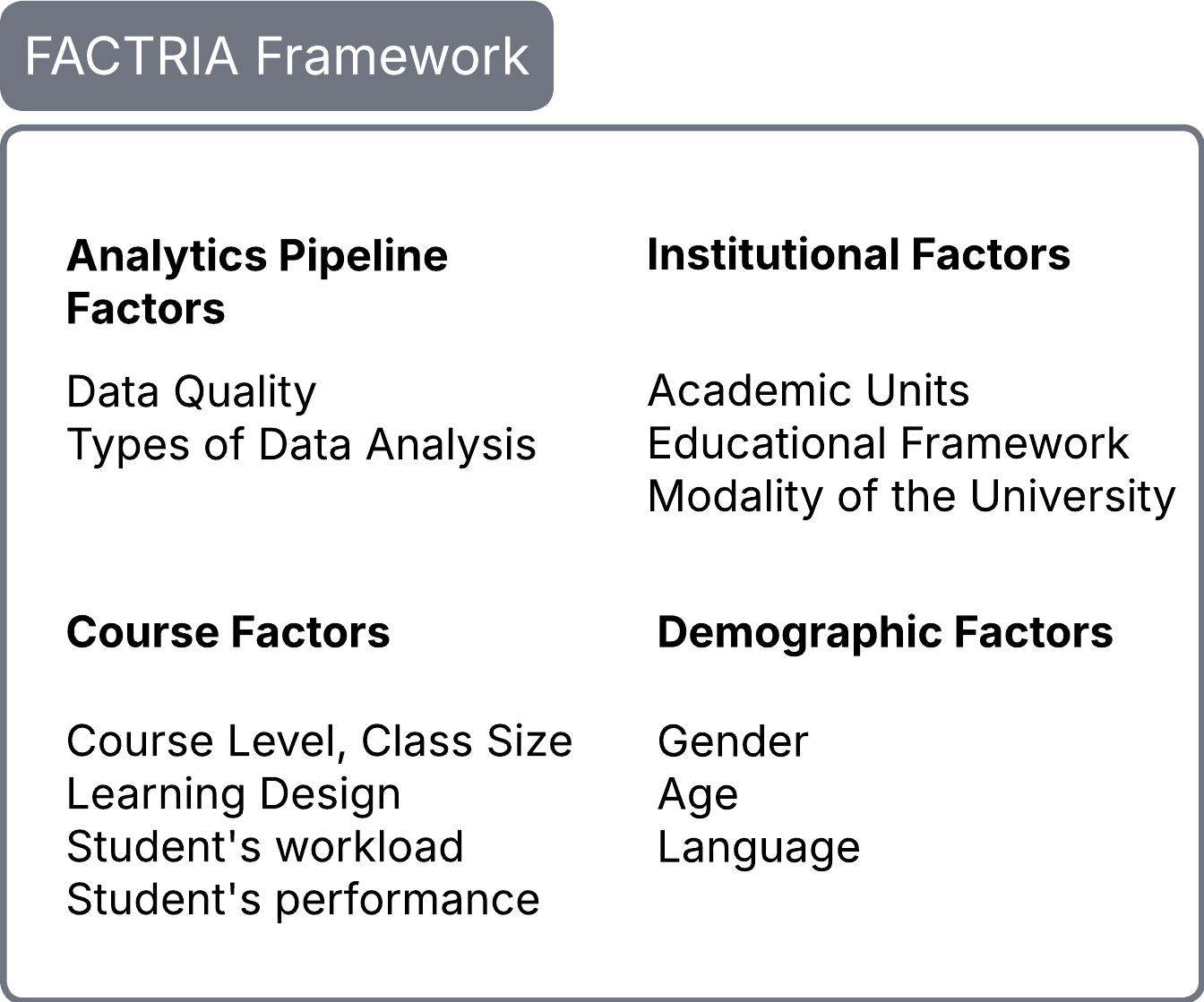}
    \caption{FACTRIA: A framework of Factors in the Analytical process and the data Context for Responsible IA.}
    \label{fig:framework}
\end{figure}


Building on prior literature, two researchers identified biasing factors that hinder the interpretation of IA and organized them into four FACTRIA categories: analytics pipeline, institutional, course, and demographic factors (see below, Appendix \ref{frameworkTable}, and Figure \ref{fig:framework}).

Existing studies emphasize not only the breadth of these factors but also the complexity of their interrelations. For example, course-level aggregations shape the indicators made available for analysis, while demographic distributions interact with these aggregations and with underlying model assumptions. Such interdependencies highlight the utility of a framework, such as FACTRIA, that foregrounds the multi-layered nature of biasing influences and supports their consideration in the design and interpretation of IA. 

Thus, aligned with the aim of this paper, the framework does not prescribe how decisions should be made. Instead, it provides a structured, evidence-informed scaffold that guides stakeholders in reflecting critically on the analytical and contextual conditions that shape their interpretations. Its purpose is to help users surface implicit assumptions, question comparability, and recognize factors that may shape interpretation—rather than to offer definitive conclusions or recommendations.

Proposing a framework, such as FACTRIA, contributes to the IA field in three key ways. First, it provides a mechanism for synthesizing and integrating currently fragmented knowledge, offering a clearer and more comprehensive account of the factors influencing IA interpretations. Second, it supports the formulation of better-grounded hypotheses and more rigorous empirical investigations by making these factors explicit. Third, it enables the development of tools, for instance, dashboards that scaffold users’ interpretation processes by drawing attention to relevant contextual or analytical considerations.

Finally, FACTRIA is a framework, i.e., it is intended as a flexible and extensible theoretical construct. As IA practices evolve and new evidence emerges, additional biasing factors may be incorporated. The framework thus serves as an adaptable foundation for advancing responsible IA.

\subsection{Analytics Pipeline Factors}

The IA pipeline represents the sequence of processes used to apply analytics to address specific problems. It includes both methodological and technical stages of data analysis, where factors that affect interpretation can emerge at any point, underscoring the importance of understanding each stage.

\begin{description}
       
    \item[Data quality] Collected data often suffers from intrinsic biases, such as missing values or inconsistencies common in educational contexts \citep{socialdatabias, Gaftandzhieva2023data-driven}. Additionally, the level of aggregation significantly impacts results \citep{radchenko2020biasesmultifaceted}; while high-level summaries protect privacy, they may obscure granular nuances such as discrimination \citep{measuringgenderbias}. \cite{Gaftandzhieva2023data-driven} highlights that higher education institutions need to consider the limitations when applying data-driven decision-making methods. Especially regarding data quality, data collected in the educational field can be inconsistent, incomplete, or unavailable.
   
    \item[Types of data analysis and visualization] This factor involves the selection of analytical techniques and visualizations, along with their underlying assumptions. For instance, selecting models that account for non-linear relationships between learning design and outcomes is crucial for accuracy \citep{ComplexAnalyticsAriel}. Regarding the assumptions underlying data analysis, \cite{measuringgenderbias} illustrates that, in specific contexts (such as the analysis of gender bias), it is essential to clearly define what constitutes gender bias to ensure the analysis is meaningful and appropriately interpreted. \cite{IAagenda} emphasizes the importance of considering context when employing evidence-based approaches, as well as the need for humans to evaluate and interpret the analysis before implementing any intervention.

\end{description}

\subsection{Institutional Factors}
Institutional factors reflect the context of Higher Education Institutions, which are crucial for fair, informed, analytics-based decisions.

\begin{description}
    \item[Academic Units] Differences may be found by analyzing structural and demographic characteristics of academic departments and centers. For instance, the Engineering center may exhibit different student outcomes than the Humanities center \citep{measuringgenderbias}. The department gender composition, which refers to the distribution of professors by gender within each department, can also have impacts that require a contextualized interpretation \citep{aragon2023gender, rosen2018correlations}.  
    \item[Educational Framework] Differences in institutional teaching practices, such as learning design guidelines, preferred teaching methodologies, assessment strategies, and constraints imposed by departmental or faculty norms, shape the context in which any institutional analysis is produced. As a result, understanding the educational framework in which the data originates is essential for making fair, contextualized interpretations of IA \citep{WiseJung2019}.
    
    \item[Modality of the University] Modality influences teaching methods, assessment strategies, student engagement, and the way data is interpreted \citep{ComplexAnalyticsAriel, usherDataActionsInstructors2021}. Different modalities generate various data types: in a traditional university, data comes from classrooms, LMS, and tools, while in an online university, data is mainly from digital activities supported by LMS. IA interpretation should account for modality differences by highlighting potential coverage gaps across courses or units based on the data sources.
\end{description}

\subsection{Course Factors}

\begin{description}
    \item[Course level, Class size] Characteristics regarding the course, such as class size, course level (e.g, first year, second year), may influence students' outcomes. For example, \cite{radchenko2020biasesmultifaceted} found biases in student satisfaction scores using a large American dataset, specifically with respect to course factors such as class size, which penalizes larger classes. In \cite{beyondbias}, the authors show that factors beyond instructors' control, such as class size and course level,  influence satisfaction scores, but also identify cases that go beyond expectations and call for a balanced, responsible interpretation.
    
    \item[Learning Design] Previous studies have shown that learning design decisions significantly affect student satisfaction and performance (e.g., \citep{thaiImpactFlippedClassroom2017} for the case of flipped classroom designs). The AL4LD framework also shows interactions between learning design factors and learning analytics \cite{hernandez2019analyticslearningdesign}. 
       
    \item[Students' workload] Research has shown that students’ perceived workload in a course correlates with other key course indicators, making it an important factor to consider—both for its potential biasing effects and for identifying cases that deviate from expected patterns—when interpreting IA \citep{beyondbias}. This factor also interacts with other elements of the educational context, particularly aspects of the learning design \citep{laurillard2013teachingasdesignscience}.
    
    \item[Student's performance] Expected or achieved grades have been shown to influence satisfaction ratings: students anticipating or receiving higher grades tend to evaluate instructors more positively \citep{badri2006identifyingpotentialbiasing, garrouste2020goodteachingandgoodgrades}. Although research has shown that the overall effect size is small and counterexamples exist, performance remains a relevant factor to consider when interpreting IA indicators.
    
\end{description}

\subsection{Demographic Factors}

Demographic factors influence IA, as unintended biases may exist, affecting data interpretation and use. Incorporating demographic awareness helps institutions reduce biases, supporting fair decision-making. 

\begin{description}

    \item[Gender] Gender may introduce potential unconscious biases in IA. \cite{measuringgenderbias} highlights the importance of considering this factor and its interactions with factors such as age and academic context, taking into account both the student's gender and the professors' gender. As with other factors, these influences are not deterministic but call for careful, responsible consideration. For example, while some studies report that female professors consistently receive lower evaluations, especially from male students \citep{mengelGenderBias, boring2017gender}, others find no significant gender-based differences in teaching quality or satisfaction when specific instructional criteria are applied \citep{Suarman2014GenderDO, arrona2020does}.

    \item[Age] Instructor age has been identified as a predictor of student evaluations, often correlating with perceived teaching effectiveness \citep{radmacher2001identifyingpredictors}. This factor also interacts with gender, with evidence showing that careful consideration should be given to analytics of female instructors who may be penalized more frequently across different age ranges compared to their male counterparts in some contexts \citep{measuringgenderbias}.

    \item[Language] Language variations and accents can shape how IA indicators are interpreted. For example, instructors with non-native accents may receive lower evaluations despite no measurable differences in student learning outcomes \citep{silaj2024impactaccentbased}. Linguistic and cultural contexts may also trigger biases linked to nationality or expected grades, which can intersect with gender and cultural prejudices \citep{genderandculturalbias, perceptionsandbiasingfactors}.
      
\end{description}

\clearpage

\section{Methods}
\label{sec:methods}

This work is grounded in the FACTRIA framework introduced in Section \ref{factria}. We used the FACTRIA framework in two ways. First, participants read a summary of the framework before the interpretation task to become familiar with the factors relevant to responsible interpretation. Second, the framework shapes the chatbot implementation, where each FACTRIA factor is encoded in the chatbot as users examine the cases with support. The remainder of this section describes the experimental design in detail.

\subsection{Study Design}

The study used a qualitative think-aloud interview design \citep{thinkAloudBook}, recording participants as they completed a structured task and verbalized their thoughts. Participants received neutral instructions to speak naturally, with minimal facilitator reminders. 

\begin{figure}[!ht]
    \centering
    \includegraphics[width=\columnwidth]{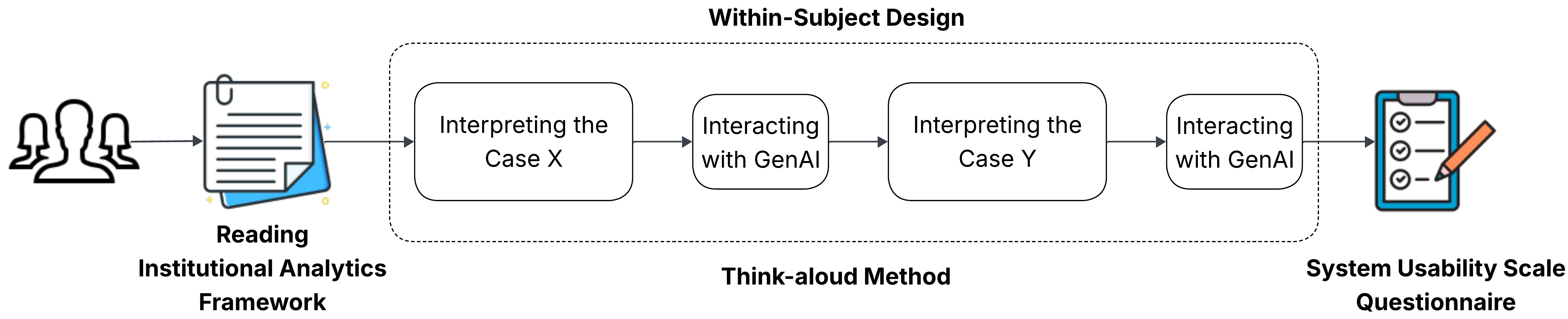}
    \caption{Study Design}
    \label{fig:studyDesign}
\end{figure}

Each participant first completed the consent form, then reviewed the FACTRIA framework (Appendix \ref{frameworkTable}) (5 minutes) before engaging in two sessions with a dashboard depicting two IA cases (Selected, from a total of four cases, for their relevance to participants’ stakeholder contexts. Further details provided in \ref{sec:participants}). For a given case, each participant (1) first interpreted an IA visualization  without the chatbot and then (2) interpreted the same visualization with FACTRIA-aware chatbot guidance (20 minutes). The order of conditions (with or without the FACTRIA-aware chatbot) was not counterbalanced, as our aim was to assess the added value of the framework and the FACTRIA-aware chatbot as an additional scaffold rather than to compare order effects. We conducted all sessions in a distraction-free environment and recorded participants' comments and on-screen dashboard activity to support a comprehensive analysis. At the end, participants completed the System Usability Scale (SUS) questionnaire \citep{usabilityScale} (Figure \ref{fig:studyDesign}). 

We conducted a cross-case analysis of all participants and IA cases to address the research question: \textit{To what extent does a chatbot leveraging FACTRIA-aware retrieval-augmented generation support stakeholders in interpreting IA responsibly?}

\subsubsection{Institutional Analytics Cases}

\label{dataViews}

The activity presented four real, context-relevant IA visualizations. We selected the cases to differ in data structure, analytical focus, and FACTRIA interpretive requirements. The visualizations originated in previous institutional studies that the participants had not previously encountered. 

\begin{figure}[hbt!]
    \centering
    \includegraphics[width=0.7\columnwidth]{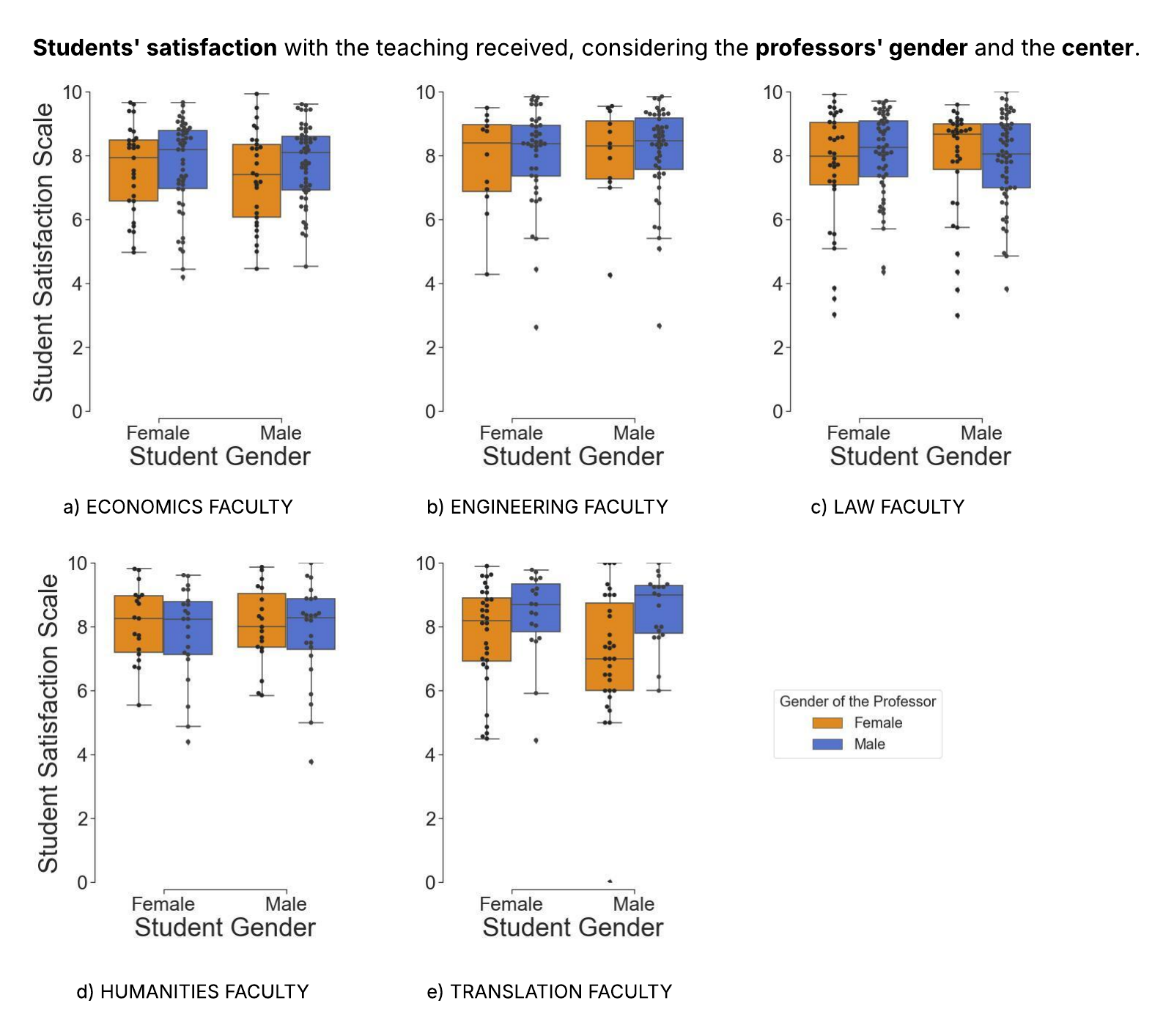}
    \caption{Case 1: Distribution of student satisfaction scores with teaching, aggregated by professor gender, student gender, and faculty. Each dot represents a professor’s mean satisfaction score \citep{measuringgenderbias}.
}
    \label{fig:case1}
\end{figure}

\begin{figure}[hbt!]
    \centering
    \includegraphics[width=\columnwidth]{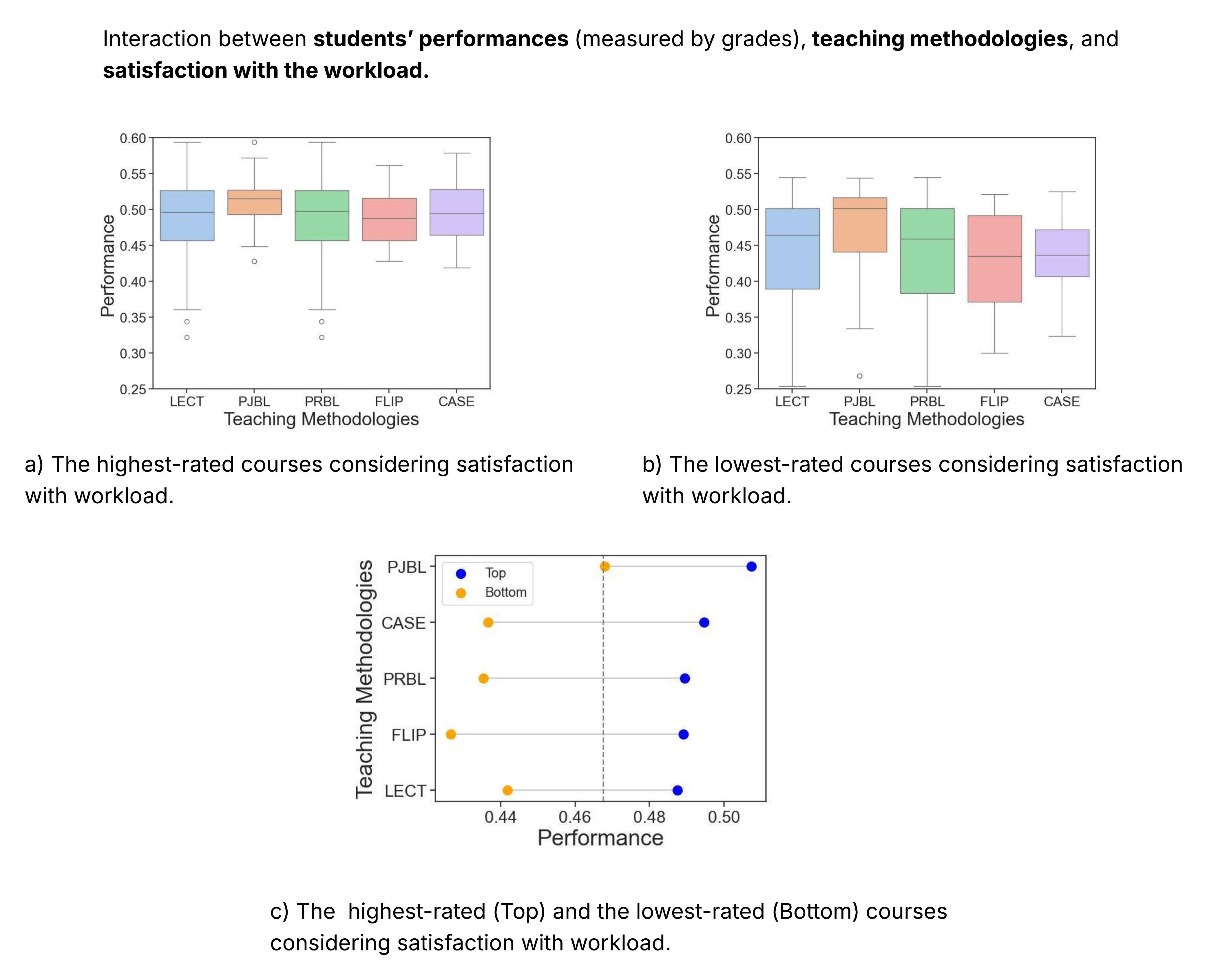}
    \caption{Case 2: Distribution of the performance (Weighted Grades) by five of the most used Active Teaching Methodologies: Lectures (LECT), Project-based learning (PJBL), Problem-based learning (PRBL), Flipped classroom (FLIP), and Case studies (CASE). Courses are ranked by workload satisfaction, and the top 20\% and bottom 20\% are highlighted based on their ratings. \citep{MarquesBehind}}
    \label{fig:case2}
\end{figure}

\begin{figure}[hbt!]
    \centering
    \includegraphics[width=\columnwidth]{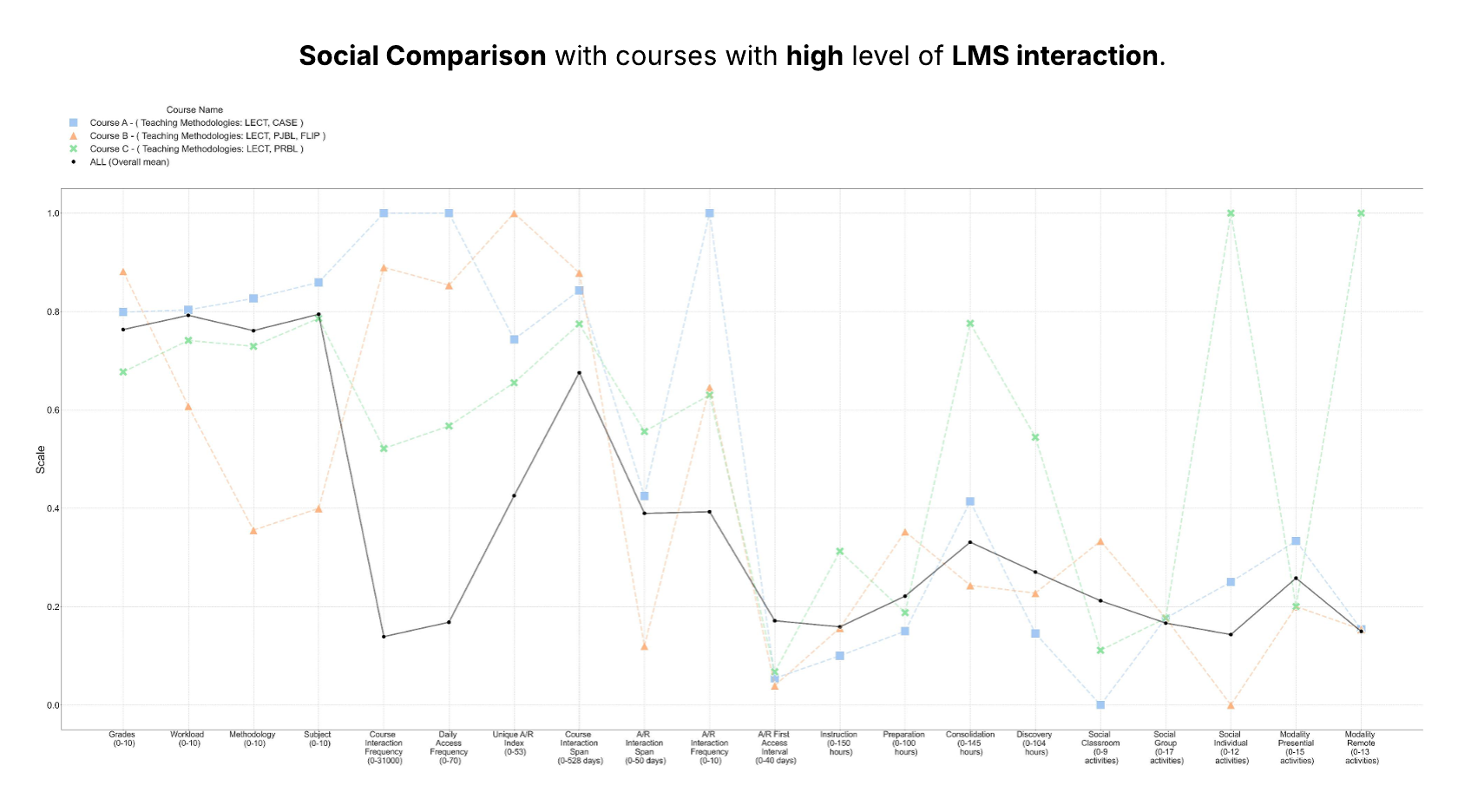}
    \caption{Case 3: Comparison of three similar high-interaction courses (A, B, C) against the institutional average on a common 0–1 scale covering 4 student outcomes (students' performance and 3 satisfaction indicators), 7 LMS usage variables at course and Task—Activity or Resource (A/R)—level, 4 planned task types, 3 social designs, and 2 task modalities. \citep{ArielMIA}}
    \label{fig:case3}
\end{figure}

\begin{figure}[!ht]
    \centering
    \includegraphics[width=0.8\columnwidth]{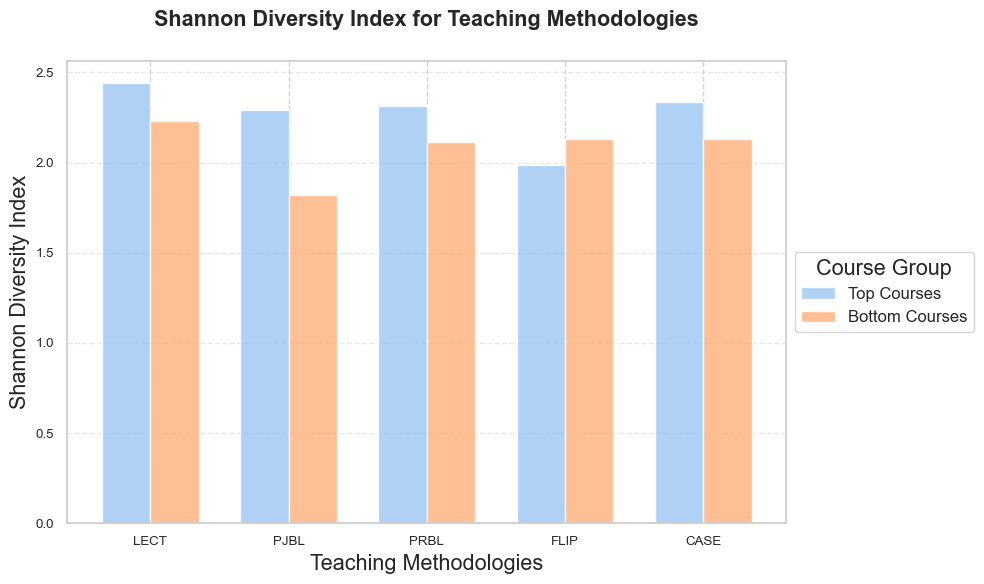}
    \caption{Case 4: Diversity of LMS course configurations using the Shannon Diversity Index (0–2.5), where higher values indicate a wider mix of activity types. Results are grouped by teaching methodology (LECT, PJBL, PRBL, FLIP, CASE) and compare courses with high versus low workload satisfaction (“Top Courses” vs. “Bottom Courses”). \citep{MarquesBehind}}
    \label{fig:case4}
\end{figure}

\begin{itemize}
    \item Case 1 presented student satisfaction with teaching, aggregated by contextual factors (see Figure \ref{fig:case1}).
    \item  Case 2 depicted the relationship among students’ academic performance (measured by grades), teaching methodologies, and perceived workload (see Figure \ref{fig:case2}). 
    \item Case 3 illustrated a social comparison across three courses characterized by higher-than-usual levels of student interaction within an LMS platform (see Figure \ref{fig:case3}). 
    \item  Case 4 displayed patterns in LMS configuration practices across different teaching methodologies (see Figure \ref{fig:case4}).
\end{itemize}

\clearpage

\subsection{Participants}
\label{sec:participants}

The participants (4 male, 7 female) included professors, academic managers, and technicians from a Spanish university. 
Table \ref{tab:profiles} summarizes these participants' anonymized profiles.

\begin{table}[htpb]
\centering
\caption{Participant Profiles. In bold are the participants and selected narratives.}
\label{tab:profiles}
\renewcommand{\arraystretch}{1.4}
\begin{tabular}{@{} >{\raggedright\arraybackslash}p{3cm} >{\raggedright\arraybackslash}p{6cm} >{\raggedright\arraybackslash}p{4.5cm} @{}}
\toprule
\textbf{Participant} & \textbf{Role and Experience} & \textbf{Use of Institutional Data} \\
\midrule
P1 (Case 3 \& 4) & LMS support specialist with 10+ years of experience. & Frequently reviews LMS data.\\
\textbf{P2} (Case 1 \& \textbf{Case 4}) & Education specialist in teaching innovation and training with 10+ years of support experience. & Checks institutional data occasionally.\\
P3 (Case 1 \& 2) & Teaching quality data support specialist with 20+ years of experience. & Frequently reviews institutional data.\\
P4 (Case 1 \& 4) & Education specialist in teaching innovation with 16+ years of support experience. & Frequently reviews institutional data.\\
P5 (Case 2 \& 3) & LMS and pedagogy support manager with 2+ years of experience. & Frequently reviews institutional data.\\
\textbf{P6} (Case 2 \& \textbf{Case 3}) & Academic leader in teaching quality and innovation, with 15+ years teaching and 4+ years management experience. & Frequently reviews institutional data.\\
\textbf{P7} (\textbf{Case 1} \& Case 2) & Academic coordinator with 25+ years teaching and 6+ years managing an undergraduate program. & Periodically reviews institutional data related to teaching quality.\\
P8 (Case 3 \& 4) & Lecturer with 15+ years of teaching experience. & Checks teaching satisfaction data occasionally.\\
P9 (Case 1 \& 4) & Lecturer with 10+ years of teaching experience. & Checks teaching satisfaction data occasionally.\\
\textbf{P10} (Case 1 \& \textbf{Case 2}) & Lecturer with 5+ years of teaching experience. & Checks teaching satisfaction data occasionally.\\
P11 (Case 1 \& 3) & Lecturer with 6+ years of teaching experience. & Checks teaching satisfaction data occasionally.\\
\bottomrule
\end{tabular}
\end{table}

\subsection{FACTRIA-Aware Chatbot Assistant}

\begin{figure}[hbt!]
    \centering
    \includegraphics[width=\columnwidth]{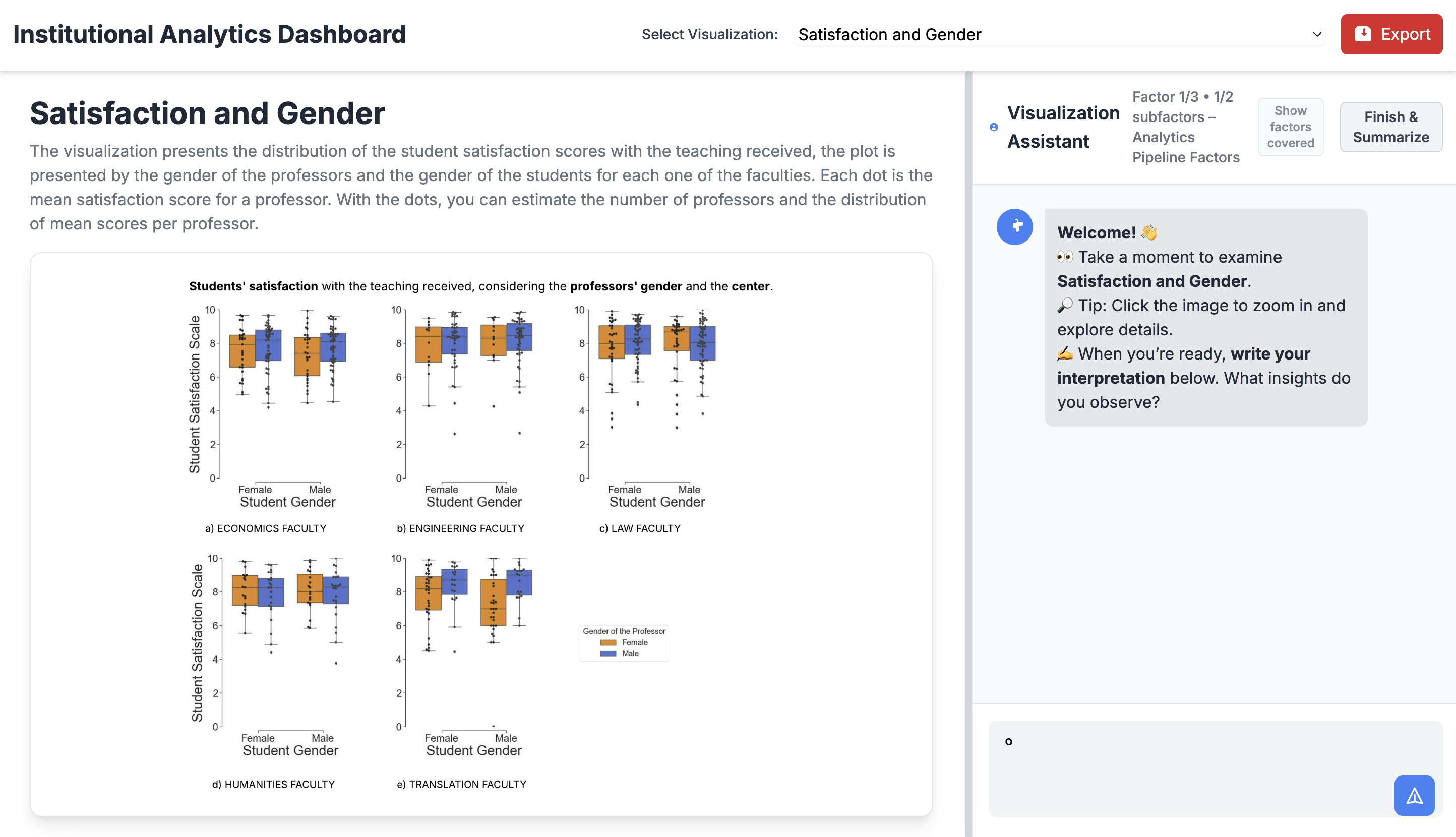}
    \caption{User interface of the chatbot}
    \label{fig:ui_screen}
\end{figure}

\begin{figure}[hbt!]
    \centering
    \includegraphics[width=0.6\columnwidth]{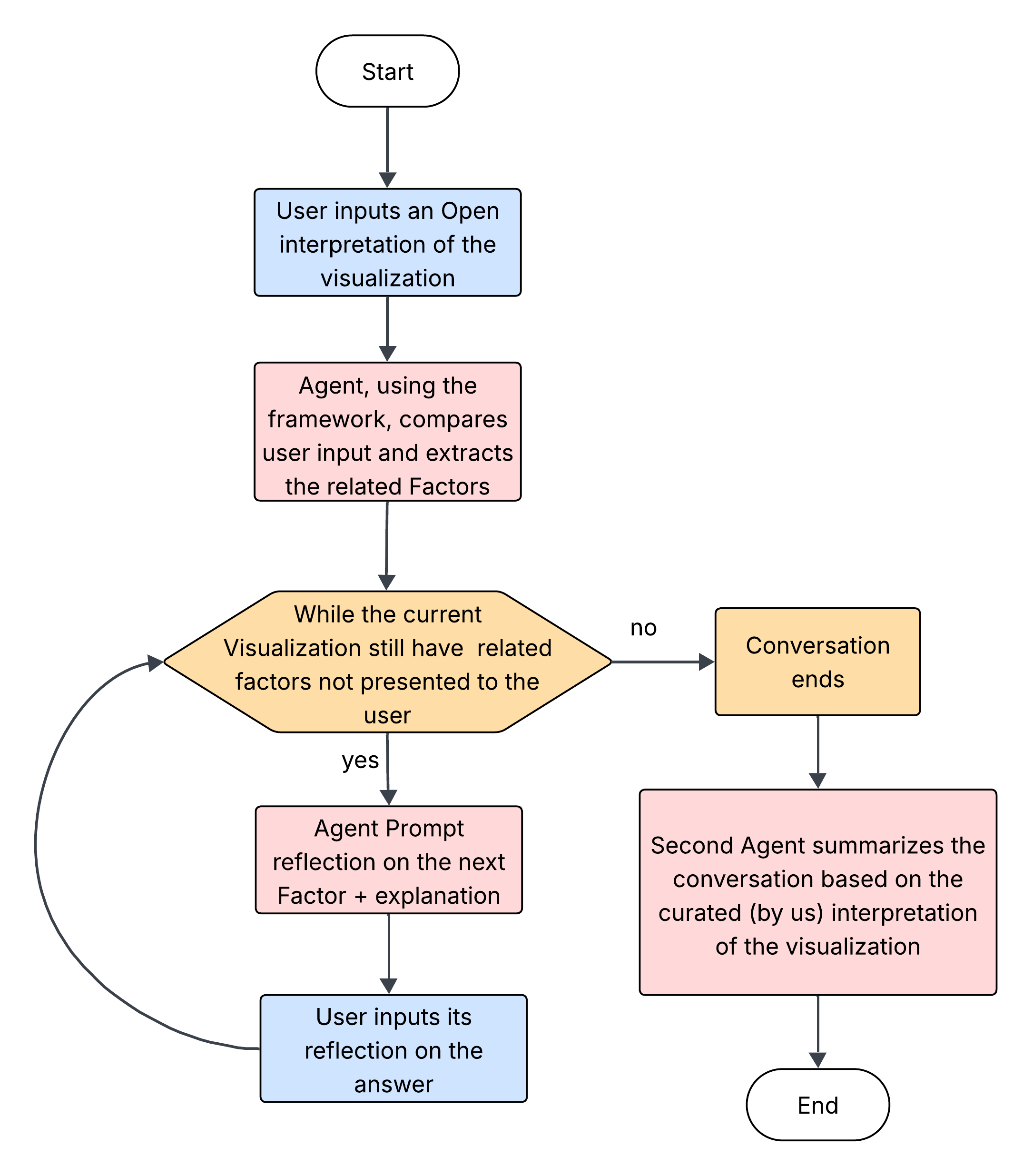}
    \caption{Flowchart describing the conversational flow between user and the chatbot}
    \label{fig:chatbot_flowchart}
\end{figure}

The FACTRIA-aware chatbot is a web-based, LLM-based system that facilitates responsible interpretation of IA, grounded in FACTRIA through retrieval-augmented generation. The dashboard interface (Figure \ref{fig:ui_screen}) integrates visualization with the chatbot, while its interaction logic (Figure \ref{fig:chatbot_flowchart}) uses a deterministic state machine to connect users’ initial interpretations to the framework and emphasize relevant factors. The system then cycles through FACTRIA factors related to the visualization, providing prompts that guide reflection towards the visible data  (for more details, see appendix \ref{techDetails}).

\subsection{Data Collection and Analysis}

The experiments were documented using the IA Dashboard, capturing both screen activity and audio.  Two researchers annotated the text data collected during participants' interactions with the FACTRIA-aware chatbot. We developed a coding scheme using a bottom-up approach (see Table \ref{codingScheme}). Two researchers first coded the transcripts independently (Cohen's k = 0.76, indicating substantial agreement). The two researchers resolved disagreements through discussion. We translated participant quotations into English where necessary and lightly edited hesitations and repetitions for readability without altering their substantive meaning.

To determine how the FACTRIA framework and the chatbot assisted users in the responsible interpretation of IA, we used transition network analysis (TNA) \citep{TNAframework} to examine interactions between users and the FACTRIA-aware chatbot, as TNA is a method for visualizing and understanding transition patterns using a probabilistic graph model.TNA was performed for each case ( see appendix \ref{AppendixTNA}) and across all cases (see Figure \ref{fig:TNAOverall}) to visualize how chatbot-based assistance facilitated responsible interpretation of IA. In addition, to understand how different participants engage with the responsible interpretation of IA with the FACTRIA-aware chatbot support, we present four narratives that reflect variation across four IA cases and distinct participant profiles.

\section{Results}
\label{results}

Through interactions with the FACTRIA-aware chatbot, participants identified the relevant factors for respective cases (3 factors for Case 1, 5 for Case 2, 5 for Case 3, and 4 for Case 4), responding to the chatbot’s prompts and ultimately producing a final summary for their analyses.

Responses to the post-activity questionnaire (see appendix \ref{tab:survey}) indicated that most participants (10 out of 11) rated the chatbot highly (4 or 5 out of 5), for providing meaningful support for contextualizing the data.In the post-survey questionnaire, Participants also reported moderate confidence when using the system, although responses varied (\textit{“I felt very confident using the system”}, $3.7 \pm 1.23$). P2 described using the system as challenging and gave its usefulness for responsible interpretation a lower rating (2 out of 5). Another participant (P8) found the system useful but described the prompts as \textit{too generic}, expressing a preference for more direct guidance.

Participants with prior experience using chatbots for similar tasks (e.g., P1, P3, P9) reported positive interactions, a pattern that is consistent with the confidence scores. However, the chatbot’s ability to provide on-demand definitions of key concepts needed to read the visualizations was especially important for less-experienced users. P6's observation that, without immediate explanations for terms such as ``weighted grade'' or ``Shannon index,'' they would have ``lost five minutes just figuring out what to do'' illustrates how on-demand definitions reduced the initial interpretive effort required to use the visualization. This is also consistent with the post-survey results, which suggest that participants generally did not feel they required substantial technical support to use the system (\textit{“I think that I would need the support of a technical person to be able to use this system”}, $2.27 \pm 1.45$), even though on-demand explanations remained important for understanding domain-specific concepts. The overall SUS score was $70.84 \pm 16.39$, which is above the conventional threshold for good usability.

\begin{figure}[hbt!]
    \centering
    \includegraphics[width=\columnwidth]{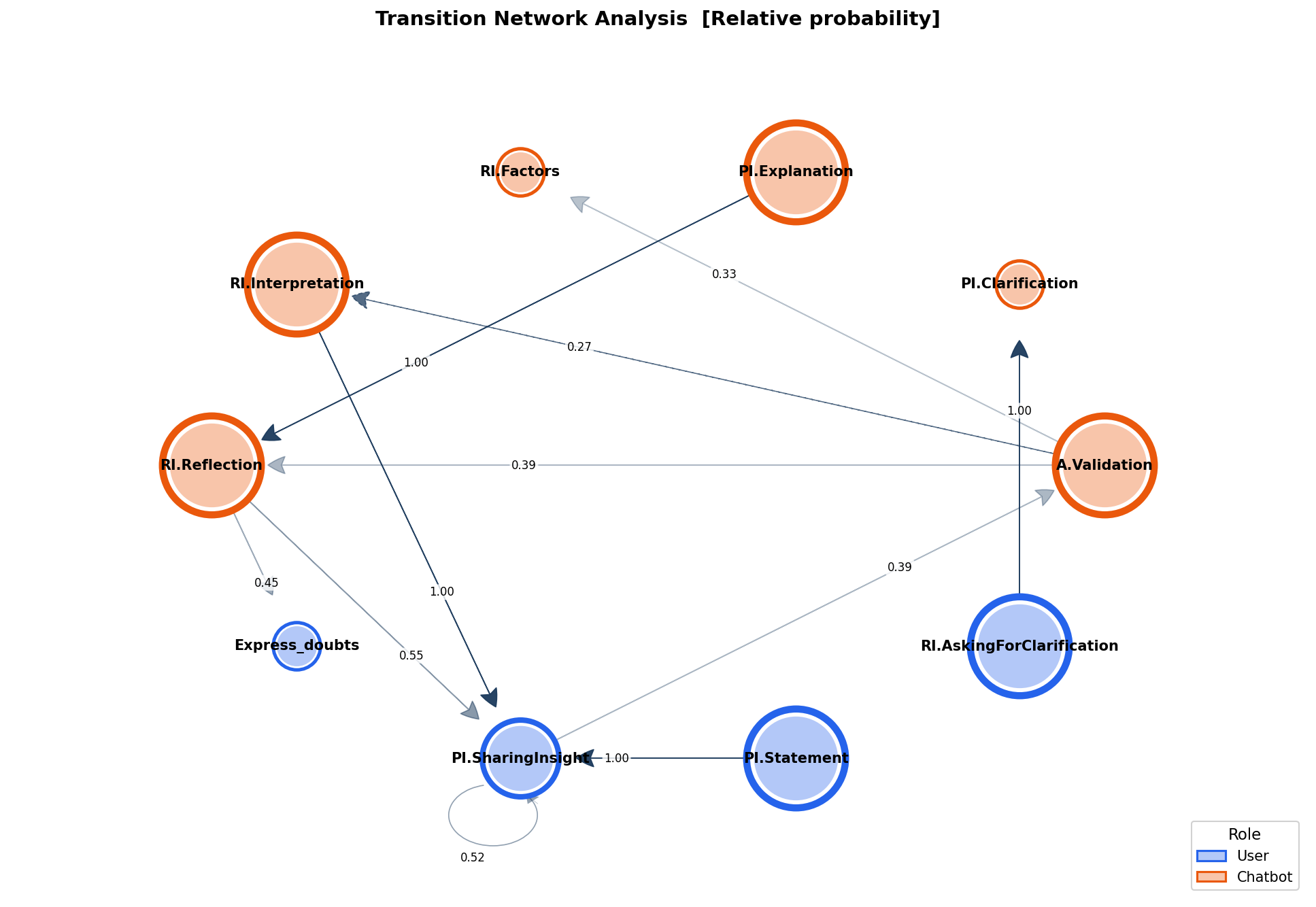}
    \caption{Transition Network Analysis considering all cases. Node size reflects the prominence of coded actions in the overall network, and edge labels represent relative transition probabilities between coded states.}
    \label{fig:TNAOverall}
\end{figure}

Figure \ref{fig:TNAOverall} shows the transition network used to examine the overall structure of chatbot–user interactions across all cases. For readability, the figure includes only transitions with relative probability above 0.25. The most prominent chatbot-coded nodes were \textit{RI.Reflection} (appeared 46 times), \textit{RI.Interpretation} (20), \textit{PI.Explanation} (26), and \textit{A.Validation} (40), while the most prominent user-coded node was \textit{PI.SharingInsight} (84) (see appendix \ref{codingScheme} for further information about the coding scheme). This distribution shows that assistant prompts and explanatory moves structured the interaction, while participants subsequently generated interpretations. Notably, the recurrence of \textit{PI.SharingInsight}, including its self-loop, shows that users often remained in an interpretive mode once they began articulating their own reasoning.

Additionally, Table \ref{tab:descriptive_results} presents descriptive statistics for the number of addressed sub-factors (without and with the FACTRIA-aware chatbot) under three scenarios. The first scenario (Case X + Case Y) includes all collected data across the study, treating each case addressed by a participant as an independent observation $(n = 22)$. We computed the paired-samples t-test, which showed a statistically significant difference between without and with the FACTRIA-aware chatbot, $p < .001$. and $Cohen′sd = 2.1530$, indicating a very large effect size. The second scenario (1° Case) restricts the analysis to the first case each participant was assigned $(n = 11)$, to control potential learning effects that could arise from participants completing a second case. In this second scenario, we also found statistically significant differences, with $p < .001$ and $Cohen′sd = 4.4947$, indicating a very large effect size. In the third scenario, we considered only the data from the second case assigned to participants, also with statistically significant differences with $n = 11$, $p < .001$. and $Cohen′sd = 1.7866$. 
In all three scenarios observed in Table \ref{tab:descriptive_results}, the use of the FACTRIA-aware chatbot increased the average number of sub-factors considered in each case.
In both cases, time constraints prevented participants from addressing all sub-factors related to the visualization using the FACTRIA-aware chatbot.

\begin{table}[htbp]
\centering

\caption{Descriptive statistics for the number of sub-factors addressed without and with the FACTRIA-aware chatbot, computed across all cases (Case X + Case Y), using only participants' first assigned case (1\textsuperscript{o} Cases) and using only participants' second assigned case (2\textsuperscript{o} Cases).}
\label{tab:descriptive_results}
\begin{tabular}{lcccccc}
\toprule
 & \multicolumn{2}{c}{\textbf{Case X + Case Y}} & \multicolumn{2}{c}{\textbf{1\textsuperscript{o} Cases}} & \multicolumn{2}{c}{\textbf{2\textsuperscript{o} Cases}} \\
\cmidrule(lr){2-3} \cmidrule(lr){4-5} \cmidrule(lr){6-7}
 & Without & With & Without & With & Without & With \\
\midrule
Average            & 1.41 & 3.45 & 1.73 & 3.55 & 1.09 & 3.36 \\
Standard deviation & 0.80 & 0.86 & 0.65 & 0.82 & 0.83 & 0.92 \\
Variance           & 0.63 & 0.74 & 0.42 & 0.67 & 0.69 & 0.85 \\
Median             & 1.50 & 4.00 & 2.00 & 4.00 & 1.00 & 4.00 \\
\bottomrule
\end{tabular}
\end{table}

Finally, we present exemplary narratives covering all four IA cases and highlighting different participant profiles (see bold in Table \ref{tab:profiles}). These narratives help illustrate how each individual makes sense of their experiences.

\subsection{Case 1}

\subsubsection{P7 – Illustrative example}

The participant’s first interpretation focused almost entirely on the clearly visible gender differences in the visualization across academic degrees. The explanation centered on the pattern that female professors appeared to receive lower evaluations than male professors. They highlighted greater discrepancies in Translation and Economics, and smaller ones in the Humanities, Law, and Engineering faculties. Much of this reasoning relied directly on the quartile distributions shown in the visualization, especially the lower quartiles: \\

\textit{“In most degrees female professors receive worse scores than male professors… the differences are more evident in the lower quartiles.” (PI.SharingInsight)}\\

At this stage, the participant interpreted the visualization as a relatively direct portrayal of gender bias, without yet invoking data quality issues, aggregation concerns, response rates, or institutional differences across academic units. 

\noindent By interacting with the FACTRIA-aware chatbot, the participant began acknowledging analytics pipeline factors—especially data quality—and, when prompted about small sample sizes, reconsidered the limitations of the aggregated indicators. \\

As the FACTRIA-aware chatbot introduced demographic factors, the participant recognized the need to consider student–professor gender interactions but remained cautious in the interpretation, recognizing the need for further inquiry: \\

\textit{“I would be cautious without knowing the sample sizes… and how the students’ gender interacts with the professors’ gender.” (PI.Caution) } \\

Although the participant did not initially reference other framework components, such as institutional differences (e.g., departmental composition), the subsequent chatbot’s prompts helped them develop a more conditional, context-aware interpretation. This shift—from straightforward readings to considering how multiple factors interact—marks a move from descriptive analysis towards responsibility-oriented inquiry. The conversation also highlighted analytics pipeline considerations, including variability in sample size and the representativeness of aggregated statistics, alongside the demographic dynamics underlying student evaluations.

\subsection{Case 2}

\subsubsection{P10 – Illustrative example}

At the start of the session, the participant focused on the chart’s most visually salient elements. Their reasoning emphasized apparent group differences without yet considering how those differences were produced. The participant initially treated the box-plot position as direct evidence of course quality: \\

\textit{“It looks like the bottom group is performing worse, because their box is lower… I guess those courses are simply weaker.” } \\

\noindent At this stage, the participant treated the visual pattern as evidence of weaker courses, without considering analytics pipeline factors such as aggregation or course factors such as class size and workload. After the FACTRIA-aware chatbot clarified that the visualization showed course-level aggregates rather than individual student data, the participant replaced the individual-level assumption with a class-size hypothesis: \\
 
\textit{“Ah, so this isn’t each student… It’s a summary of the whole course? Then maybe the variation has to do with how many students are in each class.” } \\

This marked one of the first moments in which the participant integrated analytics pipeline factors—specifically, aggregation and data transformation—into their interpretation.

When the FACTRIA-aware chatbot prompted the participant to consider course-related factors—especially workload, learning design, and class size—the participant shifted again and recognized that course structure could influence satisfaction and performance indicators: \\

\textit{“If a course has a much heavier workload or a big group of students, that could affect the scores. I didn’t think about that when I first looked at it.” } \\

\noindent This response incorporated course factors—workload and class size—as alternative explanations for the observed differences. Compared with the initial conclusion that the courses were “simply weaker,” the participant’s interpretation became more conditional and context-aware.

\subsection{Case 3}

\subsubsection{P6 – Illustrative example}

The participant’s first reading of the visualization focused on comparing courses A, B, and C with the institutional average. They noted which indicators appeared “high” or “low” relative to the black reference line and used this to infer strengths and areas for improvement. Their interpretation emphasized visible differences in methodology, subject–course interaction, and LMS-related indicators. For example, the participant described Course B as showing “lower values in methodology” relative to the reference line \textit{(PI.SharingInsight)} \\

The participant approached the visualization descriptively, using the observed differences to identify possible areas for course improvement. Although they expressed uncertainty about labels such as “Diversity index” and “social group,” their initial interpretation did not examine analytics pipeline factors related to indicator definition and data quality.

\noindent The first shift occurred when the FACTRIA-aware chatbot asked how unclear indicator definitions might influence interpretation. The participant then recognized that unclear definitions—a data quality issue within the analytics pipeline—limited the conclusions they could draw: \\
  
\textit{“I’m not fully sure about the meaning or the potential impact of this indicator.”} \textit{(Express Doubts)}\\

\noindent This response shows how an analytics pipeline concern qualified the participant’s initial comparison: interpreting the indicator values required clearer definitions of how the underlying analysis defined the measures.\\

\noindent As the FACTRIA-aware chatbot shifted to course factors, specifically learning design and workload, the participant reconsidered the interpretation regarding the meaning of the high LMS interaction levels. The participant connected interaction patterns to workload, recognizing that course design choices shape the indicators: \\

\textit{“Course A seems to have the highest interaction, and it also has the highest value of workload.” (PI.SharingInsight)} \\

\noindent This progression connected the LMS interaction indicators to course factors, particularly learning design and workload, rather than treating them as direct measures of course quality. Institutional and demographic factors did not enter the participant’s reasoning, showing the boundaries of the interpretive shift.

\subsection{Case 4}
\subsubsection{P2 – Illustrative example}

\noindent At the outset, the participant struggled to identify the purpose of the visualization, noting confusion both about the graphic and the meaning of its labels. The participant interpreted the chart as comparing satisfaction levels across teaching methodologies, rather than as showing Shannon diversity of LMS task configuration for courses with high and low workload satisfaction. This misalignment shaped their early reasoning, as they attempted to connect each methodology with its expected level of platform usage: \\

\textit{“I’m not clear about the objective of the graph… I thought it was comparing how much each methodology uses the platform.”  }\\

\noindent Based on their pedagogical understanding, the participant expected Flipped Classroom courses to use more platform resources than lectures and therefore regarded the lower diversity values as inconsistent with that expectation.

\noindent The participant also questioned the inclusion of “lectures” as a methodology and showed uncertainty about methodology labels, the Shannon Index scale, and the meaning of the top/bottom course groups. Overall, the participant grounded the interpretation in visible patterns, but an incorrect premise and unclear indicator definitions constrained it, without yet considering analytics pipeline factors related to metric definition and data representation. \\

As the FACTRIA-aware chatbot introduced factor-based prompts, the participant gradually shifted from a satisfaction-based interpretation towards the intended focus: LMS task diversity and its connection to learning design structures. The first shift occurred when the participant explicitly requested clarification of the Shannon Index \textit{(RI.AskingForClarification)}. The chatbot's explanation provided the definitional clarity needed to reconsider the visualization.

\noindent This signaled a move towards analytics pipeline factors, particularly data definitions and metric assumptions. As the FACTRIA-aware chatbot elaborated on data quality and aggregation, the participant understood that the equal weighting of all resources and activities could affect how the index represents diversity: \\

\textit{“If you apply a more complex methodology, you will need to configure more tasks in the LMS.” (PI.SharingInsight)} \\

\noindent The participant increasingly acknowledged that the portrayal of the data, rather than the index itself, was influencing their interpretation. This response connected their reasoning to analytics pipeline factors concerning data representation and types of analysis.

When the FACTRIA-aware chatbot shifted to course factors, the participant moved towards aligning the chart with their pedagogical expertise. They highlighted how task variety in Flipped Classroom should reflect multiple channels and formats to support preparation before class: \\

\textit{“Tasks should be as complete as possible, across different channels, with different application possibilities for students.” (PI.SharingInsight)} \\

\noindent The participant then connected two FACTRIA course factors, learning design and student workload, noting that methodologies like Flipped Classroom inherently require careful consideration of student workload in their design: \\

\textit{“Methodologies such as FLIP have as a key factor the consideration of the workload in their design.” (Reflection) } \\

\noindent Despite this progress, institutional factors did not emerge in this exchange, delimiting the scope of the participant’s revised interpretation. The participant's initial reading, shaped by uncertainty about the graph’s objective, treated the methodology bars as indicators of satisfaction rather than as indicators of task diversity. Through the chatbot’s guidance, the participant engaged with analytics pipeline factors (data definitions, data quality, aggregation limits) and course factors (learning design structures, workload implications, task completeness), progressively realigning their responsible interpretation with the visualization’s scope.

\section{Discussion}
\label{discussionsConclusions}

The study findings show that stakeholders often begin their interpretation of IA with narrow observations that overlook key analytical and contextual factors. Before interacting with the FACTRIA-aware chatbot, participants referenced an average of $1.50$ sub-factors ($SD = 0.80$) considering both cases, even though each case required considering multiple dimensions, from data quality and aggregation to course design and demographics. Since all participants had read the framework at the start of the experiment, familiarity alone was evidently not sufficient to prompt broad consideration in their initial interpretations. We evaluated the effect of FACTRIA-aware chatbot scaffolding under three scenarios (both cases combined, first cases only, and second cases only) and in each the chatbot surfaced sub-factors participants had not considered on their own.
This confirms concerns in the literature about the need for approaches that strengthen users’ awareness of potential biases embedded in both the analytical context and the interpretive process \citep{Williamson2022Review, socialdatabias}.

Another finding relates to how the FACTRIA-aware chatbot, design—particularly its emphasis on retrieval-augmented, framework-bound prompting—enabled participants to integrate consideration of biasing factors into their interpretations. Participants who initially relied heavily on visual cues (e.g., quartile differences in Case 1 or box-plot positions in Case 2) began, through chatbot prompts, to re-examine assumptions related to sample size, aggregation level, indicator definitions, learning-design structures, or demographic interactions. Across all participants and the four IA cases, the FACTRIA-aware chatbot supported a clear shift towards more reflective, context-aware, conditional interpretations. The qualitative data coding and transition network analysis demonstrated that the FACTRIA-aware chatbot guided the user towards a responsible interpretation by asking questions that encourage reasoning (\textit{RI.Interpretation (20 times), RI.Factors (16 times), RI.Reflection (46 times), RI.Explanation (26 times)}) instead of providing answers. This finding aligns with previous study findings showing that theory-informed structures and timely, context-aware guidance can improve users’ ability to interpret complex analytics \citep{PaulsenLindsay2024, Han2021}. This finding also complements work on AI-mediated assistance for educational dashboards \citep{VizChat, ChatLAD}, while extending that work to the institutional-level analytics context and emphasizing critical reflection rather than explanation alone.

In terms of implications for the implementation and use of IA systems, this study highlights the value of guiding stakeholders through relevant interpretive factors when working with IA. The findings suggest that frameworks such as FACTRIA can help connect dashboard indicators with the broader contextual considerations needed for sound institutional decision-making. For practice, the results point to the usefulness of embedding factor-aware guidance directly within analytics tools. Providing contextual cues at the point of interpretation can help prevent premature conclusions and promote more transparent, evidence-aligned decisions. Integrating such support into dashboards or training can strengthen institutional capacity for responsible analytics use.

\section{Limitations}
\label{limitations}

Two aspects of our study design should be considered when interpreting the findings. First, the sample included 11 participants, and each engaged with two of the four cases. While the participants represented genuine user profiles for the case under study, the experimental conditions resulted in a limited number of observations per case. Thus, essentially we treated the study as exploratory and emphasize case-level results rather than comparative. Our study’s value lies in providing a first computer-supported approach to support scaffolding in the responsible interpretation in institutional analytics.

A second limitation is that the previous knowledge about the framework was not controlled: all participants first completed the task having read the framework but without chatbot support and then completed it with FACTRIA-aware chatbot assistance. This design reflects the study’s focus on the role of the chatbot in guiding participants’ reflection, rather than on their prior knowledge of the factors included in the FACTRIA framework. We therefore examined how participants’ reflection evolved across their interactions with the chatbot in two subsequent cases, while acknowledging that their familiarity with FACTRIA was also expected to increase by the second case. Although this aspect of the experimental design was considered in the analysis and interpretation of the results, future work should examine alternative experimental conditions that allow the effects of chatbot support to be more clearly distinguished from those of increasing familiarity with the framework. 

The interpretation of the results should also account for possible automation bias, over-reliance of users on chatbot guidance, and known limitations of the think-aloud methodology, including reactivity effects and the difficulty of inferring genuine understanding from verbalized reasoning. Verbalizing interpretations may influence how participants approach the task, while differences in the detail of their responses may reflect their ability or willingness to articulate their reasoning rather than their underlying understanding. We therefore treat these data as evidence of participants’ expressed sense-making rather than as a direct measure of comprehension. Nevertheless, the think-aloud methodology strengthened the exploratory study by providing rich qualitative insights. Given these limitations, we present this work as a proof of concept. FACTRIA remains a framework that may require adaptation across cases and institutional contexts. Although the sample was relatively small, the study captures a range of authentic perspectives from a diverse set of stakeholders spanning from teachers, managers, and technicians who are directly involved in institutional data interpretation, decision-making, and taking subsequent actions at the wider institution level.  Within a single study, reaching such a diverse range of stakeholders is challenging due to practical constraints, particularly those associated with participants’ schedules and availability. Nevertheless, our study successfully generated empirical findings across these stakeholder groups, providing an important starting point for the field of institutional analytics to consider how LLMs can support the responsible interpretation of IA.

Future research should expand and refine FACTRIA by testing its adaptability across a wider range of IA cases and examining how additional sources of bias can be incorporated as institutional analytics practices and technologies evolve. Researchers should also investigate the GenAI component, given the known risks of oversimplification, occasional inaccuracies, and inconsistent alignment with institutional contexts \citep{MartinezMaldonado2020ChallengesLA, Holmes2022EthicsAIinEducation}. Applying the FACTRIA approach across a broader range of universities would further strengthen the generalizability of the findings. The findings from our study also suggest more specific directions for future research. Differences among participants indicate a need to examine how users internalize this type of scaffolding over time and whether adaptive—for example, experience-sensitive—or role-specific prompting enhances its impact. Further research could also test how different forms of explanation affect users’ sense-making when working with complex IA.

\section{Conclusion}
\label{conclusions}

This study provides an initial but meaningful response to a key gap in IA: the lack of structured support to help stakeholders consider the factors that shape responsible IA interpretation. The study offers an original contribution by synthesizing biasing factors in the FACTRIA framework that adds the needed specificity to discussions of responsible interpretation—and showing how a GenAI chatbot can operationalize this framework to scaffold users' reflective sense-making. To assess the framework’s applicability, we employed a qualitative research design involving stakeholders responsible for institutional decision-making, authentic interpretive cases, and transition network analysis. Looking ahead, further research is warranted—both to address the limitations of the present study and to investigate the new directions it opens. Collectively, this work provides a conceptual and empirical foundation for developing more transparent, fair, and contextually sound interpretive practices in IA.
\newline
\newline

\noindent \textbf{Acknowledgments}
\newline
\newline
The work was done in collaboration with UPF services (Strategy, Data and Quality Service, and the Data Protection Officer) through the UPF Learning Lab initiative.
\newline
\newline
\noindent \textbf{Funding}
\newline
\newline
This work has been supported by PID2023-146692OB-C33 and 2023-26 CEX2021-001195-M granted by MICIU/AEI/10.13039/501100011033. D. Hernández-Leo (Serra Húnter) also acknowledges the support of the AGAUR ICREA Academia programme.

\clearpage

\bibliographystyle{agsm}
\bibliography{interactcadsample}

@article{Gaftandzhieva2023data-driven,
title = {Data-driven Decision Making in Higher Education Institutions: State-of-play},
journal = {International Journal of Advanced Computer Science and Applications},
doi = {10.14569/IJACSA.2023.0140642},
year = {2023},
publisher = {The Science and Information Organization},
volume = {14},
number = {6},
author = {Silvia Gaftandzhieva and Sadiq Hussain and Slavoljub Hilcenko and Rositsa Doneva and Kirina Boykova}
}

@article{radmacher2001identifyingpredictors,
  title={Identifying significant predictors of student evaluations of faculty through hierarchical regression analysis},
  author={Radmacher, Sally A and Martin, David J},
  journal={The Journal of Psychology},
  volume={135},
  number={3},
  pages={259--268},
  year={2001},
  publisher={Taylor \& Francis}
}

@article{boring2017gender,
  title={Gender biases in student evaluations of teaching},
  author={Boring, Anne},
  journal={Journal of public economics},
  volume={145},
  pages={27--41},
  year={2017},
  publisher={Elsevier}
}

@article{silaj2024impactaccentbased,
  title={The impact of multimedia design and the accent of the instructor on student learning and evaluations of teaching},
  author={Silaj, Katie M and Frangiyyeh, Antonia and Paquette-Smith, Melissa},
  journal={Applied Cognitive Psychology},
  volume={38},
  number={1},
  pages={e4143},
  year={2024},
  publisher={Wiley Online Library}
}

@article{socialdatabias,
  title={Social data: Biases, methodological pitfalls, and ethical boundaries},
  author={Olteanu, Alexandra and Castillo, Carlos and Diaz, Fernando and K{\i}c{\i}man, Emre},
  journal={Frontiers in big data},
  volume={2},
  pages={13},
  year={2019},
  publisher={Frontiers Media SA}
}

@article{genderandculturalbias,
  title={Gender and cultural bias in student evaluations: Why representation matters},
  author={Fan, Yanan and Shepherd, Laura J and Slavich, Eve and Waters, David and Stone, M and Abel, Rachel and Johnston, Emma L},
  journal={PloS one},
  volume={14},
  number={2},
  pages={e0209749},
  year={2019},
  publisher={Public Library of Science San Francisco, CA USA}
}

@article{arrona2020does,
  title={Does professors' gender impact how students evaluate their teaching and the recommendations for the best professor?},
  author={Arrona-Palacios, Arturo and Okoye, Kingsley and Camacho-Zu{\~n}iga, Claudia and Hammout, Nisrine and Luttmann-Nakamura, Emilia and Hosseini, Samira and Escamilla, Jose},
  journal={Heliyon},
  volume={6},
  number={10},
  pages={e05313},
  year={2020},
  publisher={Elsevier}
}

@article{RIENTIES2016333across151modules,
title = {The impact of learning design on student behaviour, satisfaction and performance: A cross-institutional comparison across 151 modules},
journal = {Computers in Human Behavior},
volume = {60},
pages = {333-341},
year = {2016},
issn = {0747-5632},
author = {Bart Rienties and Lisette Toetenel}
}

@article{rosen2018correlations,
  title={Correlations, trends and potential biases among publicly accessible web-based student evaluations of teaching: a large-scale study of RateMyProfessors. com data},
  author={Rosen, Andrew S},
  journal={Assessment \& Evaluation in Higher Education},
  volume={43},
  number={1},
  pages={31--44},
  year={2018},
  publisher={Taylor \& Francis}
}

@article{Suarman2014GenderDO,
  title={Gender Differences on Students Satisfaction: The Role of Teaching Quality in Higher Education},
  author={Suarman},
  journal={Middle-East Journal of Scientific Research},
  volume={21},
  number={9},
  pages={1434-1441},
  year={2014}
}

@article{heffernan2023abusive,
  title={Abusive comments in student evaluations of courses and teaching: the attacks women and marginalised academics endure},
  author={Heffernan, Troy},
  journal={Higher Education},
  volume={85},
  number={1},
  pages={225--239},
  year={2023},
  publisher={Springer}
}

@article{li2016modelling,
  title={Modelling and managing learner satisfaction: Use of learner feedback to enhance blended and online learning experience},
  author={Li, Nai and Marsh, Vicky and Rienties, Bart},
  journal={Decision Sciences Journal of Innovative Education},
  volume={14},
  number={2},
  pages={216--242},
  year={2016},
  publisher={Wiley Online Library}
}

@article{garrouste2020goodteachingandgoodgrades,
  title={Good teaching and good grades. Can you buy pedagogy?},
  author={Garrouste, Manon and Le Saout, Ronan},
  journal={Annals of Economics and Statistics},
  volume={139},
  pages={29--60},
  year={2020},
  publisher={JSTOR}
}

@article{badri2006identifyingpotentialbiasing,
  title={Identifying potential biasing variables in student evaluation of teaching in a newly accredited business program in the UAE},
  author={Badri, Masood A and Abdulla, Mohamed and Kamali, Mohammed A and Dodeen, Hamzeh},
  journal={International Journal of Educational Management},
  volume={20},
  number={1},
  pages={43--59},
  year={2006},
  publisher={Emerald Group Publishing Limited}
}

@article{radchenko2020biasesmultifaceted,
title={Biases in Student Evaluations of Teaching: An American Case Study},
  author={Radchenko, Natalia},
  year={2020},
  journal={IZA Discussion Paper},
}

@article{hernandez2019analyticslearningdesign,
  title={Analytics for learning design: A layered framework and tools},
  author={Hern{\'a}ndez-Leo, Davinia and Martinez-Maldonado, Roberto and Pardo, Abelardo and Mu{\~n}oz-Crist{\'o}bal, Juan A and Rodr{\'\i}guez-Triana, Mar{\'\i}a J},
  journal={British Journal of Educational Technology},
  volume={50},
  number={1},
  pages={139--152},
  year={2019},
  publisher={Wiley Online Library}
}

@book{laurillard2013teachingasdesignscience,
  title={Teaching as a design science: Building pedagogical patterns for learning and technology},
  author={Laurillard, Diana},
  year={2013},
  publisher={Routledge}
}

@article{professorAgeandGender,
  title={Do professor’s age and gender matter or do students give higher value to professors’ experience?},
  author={Flegl, Martin and Andrade Rosas, Luis Antonio},
  journal={Quality Assurance in Education},
  volume={27},
  number={4},
  pages={511--532},
  year={2019},
  publisher={Emerald Publishing Limited}
}

@article{mengelGenderBias,
    author = {Mengel, Friederike and Sauermann, Jan and Zölitz, Ulf},
    title = "{Gender Bias in Teaching Evaluations}",
    journal = {Journal of the European Economic Association},
    volume = {17},
    number = {2},
    pages = {535-566},
    year = {2018},
    month = {02},
    issn = {1542-4766}
}

@article{AreTheyRelatedtoWhatStudentsLearn,
author = {Dennis E. Clayson},
title ={Student Evaluations of Teaching: Are They Related to What Students Learn?: A Meta-Analysis and Review of the Literature},

journal = {Journal of Marketing Education},
volume = {31},
number = {1},
pages = {16-30},
year = {2009},

}

@article{beltran2022surviving,
  title={Surviving and thriving: How changes in teaching modalities influenced student satisfaction before, during and after COVID-19},
  author={Ortiz-Beltr{\'a}n, Ariel  and Hern{\'a}ndez-Leo, Davinia and Amarasinghe, Ishari},
  journal={Australasian Journal of Educational Technology},
  pages={72--88},
  year={2022}
}

@article{perceptionsandbiasingfactors,
  title={Student evaluations of teaching: perceptions and biasing factors},
  author={Al-Issa, Ahmad and Sulieman, Hana},
  journal={Quality Assurance in Education},
  volume={15},
  number={3},
  pages={302--317},
  year={2007},
  publisher={Emerald Group Publishing Limited}
}

@article{aragon2023gender,
  title={Gender bias in teaching evaluations: the causal role of department gender composition},
  author={Arag{\'o}n, Oriana R and Pietri, Evava S and Powell, Brian A},
  journal={Proceedings of the National Academy of Sciences},
  volume={120},
  number={4},
  year={2023},
  publisher={National Acad Sciences}
}

@article{paz2019academic,
  title={Academic analytics: a systematic review of literature},
  author={Paz, F{\'a}bio Josende and Cazella, Silvio Cesar},
  journal={International Journal of Development Research},
  volume={9},
  number={11},
  pages={31710--31716},
  year={2019}
}

@article{yanosky2015analyticslandscape,
  title={The analytics landscape in higher education},
  author={Yanosky, Ronald and Arroway, Pam},
  journal={Louisville, CO: Educase Center for Analysis and Research},
  year={2015},
}

@article{LAforLDreview,
author = {Mangaroska, Katerina and Giannakos, Michail},
title = {Learning Analytics for Learning Design: A Systematic Literature Review of Analytics-Driven Design to Enhance Learning},
year = {2019},
issue_date = {Oct.-Dec. 2019},
publisher = {IEEE Computer Society Press},
address = {Washington, DC, USA},
volume = {12},
number = {4},
issn = {1939-1382},
journal = {IEEE Trans. Learn. Technol.},
month = {oct},
pages = {516–534},
numpages = {19}
}

@article{whatarewemeasuring,
  title={University performance evaluations: what are we really measuring?},
  author={Bedggood, Rowan E and Donovan, Jerome D},
  journal={Studies in Higher Education},
  volume={37},
  number={7},
  pages={825--842},
  year={2012},
  publisher={Taylor \& Francis}
}

@article{thaiImpactFlippedClassroom2017,
  title = {The Impact of a Flipped Classroom Design on Learning Performance in Higher Education: {{Looking}} for the Best ``Blend'' of Lectures and Guiding Questions with Feedback},
  shorttitle = {The Impact of a Flipped Classroom Design on Learning Performance in Higher Education},
  author = {Thai, Ngoc Thuy Thi and De Wever, Bram and Valcke, Martin},
  year = {2017},
  month = apr,
  journal = {Computers \& Education},
  volume = {107},
  pages = {113--126},
  issn = {0360-1315}
}

@article{measuringgenderbias,
author = {Francielle Marques and Davinia Hernández-Leo and Carlos Castillo},
title = {Measuring gender bias in student satisfaction in higher education: a cross-department study},
journal = {Cogent Education},
volume = {11},
number = {1},
pages = {2375183},
year = {2024},
publisher = {Cogent OA},
}

@article{beyondbias,
  title={Beyond bias in student satisfaction surveys: exploring the role of grades and satisfaction with the learning design},
  author={Marques, Francielle and Hern{\'a}ndez-Leo, Davinia and Castillo, Carlos},
  journal={Journal of New Approaches in Educational Research},
  volume={14},
  number={1},
  pages={1--13},
  year={2025},
  publisher={Springer}
}

@article{IAagenda,
  title={Toward an Institutional Analytics Agenda for Addressing Student Dropout in Higher Education: An Academic Stakeholders' Perspective.},
  author={De Silva, Liyanachchi Mahesha Harshani and Chounta, Irene-Angelica and Rodr{\'\i}guez-Triana, Mar{\'\i}a Jes{\'u}s and Roa, Eric Roldan and Gramberg, Anna and Valk, Aune},
  journal={Journal of Learning Analytics},
  volume={9},
  number={2},
  pages={179--201},
  year={2022},
  publisher={ERIC}
}

@inproceedings{educationaldatamining,
author = {Llaur\'{o}, Alba and Fonseca, David and Villegas, Eva and Al\'{a}ez, Marian and Romero, Susana},
title = {Educational data mining application for improving the academic tutorial sessions, and the reduction of early dropout in undergraduate students},
year = {2021},
isbn = {9781450390668},
publisher = {Association for Computing Machinery},
address = {New York, NY, USA},
booktitle = {Ninth International Conference on Technological Ecosystems for Enhancing Multiculturality (TEEM'21)},
pages = {212–218},
numpages = {7},
location = {Barcelona, Spain},
series = {TEEM'21}
}

@article{HELACMM,
  title={{HELA-CMM}: capability maturity model for adoption of learning analytics in higher education},
  author={{\v{S}}imi{\'c}, Diana and Begi{\v{c}}evi{\'c} Re{\dj}ep, Nina and Rako, Sabina and Kadoi{\'c}, Nikola and Van Petegem, Wim and Rienties, Bart and Cabrera Lanzo, Nati and Eichhorn, Michael and Gu{\`a}rdia, Lourdes and Ku{\v{c}}ina Softi{\'c}, Sandra and Alexander Tillmann},
  journal={International journal of educational technology in higher education},
  volume={22},
  number={1},
  pages={25},
  year={2025},
  publisher={Springer}
}

@article{modernFramework,
author={Weil, Dave and Kendall, Casey and Snyder, Rob},
year={2023},
month={Feb 14},
title={A Modern Framework for Institutional Analytics},
journal={EDUCAUSE Review (Online)},
language={English}
}

@InProceedings{ComplexAnalyticsAriel,
author="Ortiz-Beltr{\'a}n, Ariel
and Marques, Francielle
and Hern{\'a}ndez-Leo, Davinia",
editor="Tammets, Kairit
and Sosnovsky, Sergey
and Ferreira Mello, Rafael
and Pishtari, Gerti
and Nazaretsky, Tanya",
title="Exploring the Complex Analytics Interplay of LMS Design, Usage, Academic Outcomes, and Perceived Workload: A Case Study",
booktitle="Two Decades of TEL. From Lessons Learnt to Challenges Ahead",
year="2025",
publisher="Springer Nature Switzerland",
address="Cham",
pages="212--217",
isbn="978-3-032-03873-9"
}

@inproceedings{DELICATEframework,
  title={Privacy and analytics: it's a DELICATE issue a checklist for trusted learning analytics},
  author={Drachsler, Hendrik and Greller, Wolfgang},
  booktitle={Proceedings of the sixth international conference on learning analytics \& knowledge},
  pages={89--98},
  year={2016}
}

@article{usabilityScale,
  title={SUS-A quick and dirty usability scale},
  author={Brooke, John and others},
  journal={Usability evaluation in industry},
  volume={189},
  number={194},
  pages={4--7},
  year={1996},
  publisher={London, England}
}

@article{WiseJung2019,
  author    = {Alyssa F. Wise and Yeonji Jung},
  title     = {Teaching with Analytics: Towards a Situated Model of Instructor Decision-Making},
  journal   = {Journal of Learning Analytics},
  year      = {2019},
  volume    = {6},
  number    = {2},
  pages     = {53--69}
}

@techreport{Rienties2016,
  author      = {Bart Rienties and Avinash Boroowa and Simon Cross and C. Kubiak and Kevin Mayles and Sam Murphy},
  title       = {Analytics4Action Evaluation Framework: A Review of Evidence-Based Learning Analytics Interventions at the Open University UK},
  institution = {The Open University},
  year        = {2016}
}

@techreport{Rienties2018,
  author      = {Bart Rienties and Avinash Boroowa and Simon Cross and C. Kubiak and Kevin Mayles and Sam Murphy},
  title       = {Analytics4Action Evaluation Framework---A Review of Evidence-Based Learning Analytics Interventions at the Open University UK},
  institution = {The Open University},
  year        = {2018}
}

@inproceedings{AtkissonWiley2011,
  author    = {Michael Atkisson and David A. Wiley},
  title     = {Learning Analytics as Interpretive Practice: Applying Westerman to Educational Intervention},
  booktitle = {Proceedings of the 1st International Conference on Learning Analytics \& Knowledge (LAK)},
  year      = {2011},
  pages     = {22--23},
}

@article{Lem2014,
  author    = {Lem, Stephanie and Onghena, Patrick and Verschaffel, Lieven and Van Dooren, Wim},
  title     = {Experts’ Misinterpretation of Box Plots – a Dual Processing Approach},
  journal   = {Psychologica Belgica},
  year      = {2014},
  volume    = {54},
  number    = {4},
  pages     = {395--405}
}

@article{Pozdniakov2025,
  author    = {Pozdniakov, Stanislav and Martínez-Maldonado, Roberto and Tsai, Yi-Shan and Echeverría, Vanessa and Swiecki, Zachary and Ga{\v{s}}evi{\'c}, Dragan},
  title     = {Investigating the Effect of Visualization Literacy and Guidance on Teachers’ Dashboard Interpretation},
  journal   = {Journal of Learning Analytics},
  year      = {2025},
  volume    = {12},
  number    = {1},
  pages     = {367--390}
}

@article{Susnjak2022,
  author    = {Susnjak, Teo and Ramaswami, Gomathy Suganya and Mathrani, Anuradha},
  title     = {Learning analytics dashboard: A tool for providing actionable insights to learners},
  journal   = {International Journal of Educational Technology in Higher Education},
  year      = {2022},
  volume    = {19},
  pages     = {12},
}

@article{usherDataActionsInstructors2021,
  title = {From Data to Actions: {{Instructors}}' Decision Making Based on Learners' Data in Online Emergency Remote Teaching},
  shorttitle = {From Data to Actions},
  author = {Usher, Maya and Hershkovitz, Arnon and Forkosh‐Baruch, Alona},
  year = {2021},
  journaltitle = {British Journal of Educational Technology},
  shortjournal = {Brit J Educational Tech},
  volume = {52},
  number = {4},
  pages = {1338--1356},
  issn = {0007-1013, 1467-8535},
  urldate = {2025-11-15},
  langid = {english}
}

@article{PaulsenLindsay2024,
  author       = {Paulsen, Christine and Lindsay, Jenni},
  title        = {Learning analytics dashboards are increasingly becoming about learning and not just analytics: A systematic review},
  journal      = {Education and Information Technologies},
  year         = {2024}
}

@article{Han2021,
  author       = {Han, Jing and Chen, Fang and Kitto, Kirsty},
  title        = {Learning analytics dashboards for adaptive support in face-to-face collaborative argumentation},
  journal      = {Computers \& Education},
  year         = {2021},
  volume       = {163},
  pages        = {104099}
}

@article{Jones2019LAHE,
  author    = {Jones, K. M. L.},
  title     = {Learning analytics and higher education: A proposed model for adoption},
  journal   = {International Journal of Educational Technology in Higher Education},
  volume    = {16},
  number    = {1},
  pages     = {1--16},
  year      = {2019}
}

@inproceedings{ChatLAD,
author = {Yang, Chao and Wang, Deliang and Chen, Gaowei},
title = {Chat-LAD: Enhancing Teacher Understanding of Learning Analytics Dashboard with AI-Empowered Explanations},
year = {2025},
isbn = {9798400712913},
publisher = {Association for Computing Machinery},
address = {New York, NY, USA},
booktitle = {Proceedings of the Twelfth ACM Conference on Learning @ Scale},
pages = {197–201},
numpages = {5},
location = {Palermo, Italy},
series = {L@S '25}
}

@inproceedings{VizChat,
  title={VizChat: enhancing learning analytics dashboards with contextualised explanations using multimodal generative AI chatbots},
  author={Yan, Lixiang and Zhao, Linxuan and Echeverria, Vanessa and Jin, Yueqiao and Alfredo, Riordan and Li, Xinyu and Ga{\v{s}}evi’c, Dragan and Martinez-Maldonado, Roberto},
  booktitle={International conference on artificial intelligence in education},
  pages={180--193},
  year={2024},
  organization={Springer}
}

@article{Holmes2022EthicsAIinEducation,
  author    = {Holmes, Wayne and Bialik, Maya and Fadel, Charles},
  title     = {Ethics of AI in Education: Towards a Community-Wide Framework},
  journal   = {British Journal of Educational Technology},
  volume    = {53},
  number    = {2},
  pages     = {550--566},
  year      = {2022}
}

@inproceedings{MartinezMaldonado2020ChallengesLA,
  author    = {Martínez-Maldonado, Roberto},
  title     = {Human-Centred Learning Analytics: Four Challenges in Dashboard Design},
  booktitle = {Proceedings of the 10th International Conference on Learning Analytics \& Knowledge (LAK ’20)},
  pages     = {100--105},
  year      = {2020},
}

@inproceedings{Williamson2022Review,
doi = {10.1145/3506860.3506900},
author = {Williamson, Kimberly and Kizilcec, Rene},
title = {A Review of Learning Analytics Dashboard Research in Higher Education: Implications for Justice, Equity, Diversity, and Inclusion},
year = {2022},
isbn = {9781450395731},
publisher = {Association for Computing Machinery},
address = {New York, NY, USA},
booktitle = {LAK22: 12th International Learning Analytics and Knowledge Conference},
pages = {260–270},
numpages = {11},
location = {Online, USA},
series = {LAK22}
}

@article{analyticsHEI,
  title={Analytics in Higher Education Benefits, Barriers, Progress, and Recommendations  (Research Report)},
  author={Bichsel, Jacqueline},
  journal={Louisville, CO: EDUCAUSE Center for Applied Research},
  year={2012},
  url={https://library.educause.edu/~/media/files/library/2012/6/ers1207.pdf}
}

@book{thinkAloudBook,
  title={Using think-aloud interviews and cognitive labs in educational research},
  author={Leighton, Jacqueline P},
  year={2017},
  publisher={Oxford University Press}
}

@inproceedings{TNAframework,
  title={Transition network analysis: A novel framework for modeling, visualizing, and identifying the temporal patterns of learners and learning processes},
  author={Saqr, Mohammed and L{\'o}pez-Pernas, Sonsoles and T{\"o}rm{\"a}nen, Tiina and Kaliisa, Rogers and Misiejuk, Kamila and Tikka, Santtu},
  booktitle={Proceedings of the 15th international learning analytics and knowledge conference},
  pages={351--361},
  year={2025}
}

@article{shi2026llmreview,
  title = {Large Language Models in Education: A Systematic Review of Empirical Applications, Benefits, and Challenges},
  author = {Shi, Yuhong and Yu, Kun and Dong, Yifei and Chen, Fang},
  year = 2026,
  journal = {Computers and Education: Artificial Intelligence},
  volume = {10},
  pages = {100529},
  issn = {2666-920X},
  doi = {10.1016/j.caeai.2025.100529}
}

@article{calvo2019improving,
  title={Improving teacher game learning analytics dashboards through ad-hoc development},
  author={Calvo Morata, Antonio and Alonso Fern{\'a}ndez, Cristina and P{\'e}rez Colado, Iv{\'a}n Jos{\'e} and Freire Mor{\'a}n, Manuel and Mart{\'\i}nez Ortiz, Iv{\'a}n and Fern{\'a}ndez Manj{\'o}n, Baltasar},
  year={2019}
}

@inproceedings{SHEILA,
author = {Tsai, Yi-Shan and Moreno-Marcos, Pedro Manuel and Tammets, Kairit and Kollom, Kaire and Ga{\v s}evi{\'c}, Dragan},
title = {SHEILA policy framework: informing institutional strategies and policy processes of learning analytics},
year = {2018},
isbn = {9781450364003},
publisher = {Association for Computing Machinery},
address = {New York, NY, USA},
url = {https://doi.org/10.1145/3170358.3170367},
doi = {10.1145/3170358.3170367},
booktitle = {Proceedings of the 8th International Conference on Learning Analytics and Knowledge},
pages = {320–329},
numpages = {10},
location = {Sydney, New South Wales, Australia},
series = {LAK '18}
}

@inproceedings{CALVI,
author = {Ge, Lily W. and Cui, Yuan and Kay, Matthew},
title = {CALVI: Critical Thinking Assessment for Literacy in Visualizations},
year = {2023},
isbn = {9781450394215},
publisher = {Association for Computing Machinery},
address = {New York, NY, USA},
url = {https://doi.org/10.1145/3544548.3581406},
doi = {10.1145/3544548.3581406},
booktitle = {Proceedings of the 2023 CHI Conference on Human Factors in Computing Systems},
articleno = {815},
numpages = {18},
location = {Hamburg, Germany},
series = {CHI '23}
}

@unpublished{ArielMIA,
  title={Multidimensional Institutional Analytics (MIA): Turning Institutional Data into Design-Oriented Hypotheses.},
  author={Ortiz-Beltrán, Ariel and Marques, Francielle and  Hernandez-Leo, Davinia},
  note   = {Manuscript under review},

}

@unpublished{MarquesBehind,
 title={ Behind the Extremes: Leveraging Institutional Analytics to Bridge Learning Design and Student Outcomes.},
  author={Marques, Francielle and Ortiz-Beltrán, Ariel and Hernandez-Leo, Davinia},
  note   = {Manuscript under review},

}

\clearpage

\appendix

\newgeometry{margin=1.5cm}


\begin{landscape}
{\normalsize
\renewcommand{\arraystretch}{1.25}
\setlength{\tabcolsep}{4pt}
\setlength{\LTcapwidth}{\linewidth}
\begin{longtable}{L{4.2cm} L{21.6cm}}
\caption{Summary of studies that influenced the development of FACTRIA framework}
\label{tab:factors} \\
\toprule
\rowcolor{gray!15}
\textbf{Reference} & \textbf{Description} \\
\midrule
\endfirsthead
\toprule
\rowcolor{gray!15}
\textbf{Reference} & \textbf{Description} \\
\midrule
\endhead
\bottomrule
\endfoot

\cite{radchenko2020biasesmultifaceted} & Described the aspects that may influence student satisfaction and the effects that result in biases. The author argues that various factors, such as gender and course level, influence data collection on student satisfaction. \\
\cmidrule{1-2}
\cite{measuringgenderbias} & Described the importance of considering reliability to analyze the data collection. Also investigated gender biases by studying different factors such as age, level of the course, and academic centers. \\
\cmidrule{1-2}
\cite{AreTheyRelatedtoWhatStudentsLearn} & Examined previous research on the connection between learning and student satisfaction. The findings indicated that, on average, there is only a weak relationship between the two variables, depending on factors such as context, discipline, course type, and instructor. \\
\cmidrule{1-2}
\cite{beltran2022surviving} & Evaluated how changes in teaching modalities influenced student satisfaction scores before, during, and after COVID-19. \\
\cmidrule{1-2}
\cite{rosen2018correlations} & Investigated gender and the impact of teaching evaluations across various academic departments and disciplines. \\
\cmidrule{1-2}
\cite{aragon2023gender} & Investigated the impact of department gender composition on student satisfaction scores. \\
\cmidrule{1-2}
\cite{RIENTIES2016333across151modules} & Studied the impact of learning design to understand student behavior and performance. \\
\cmidrule{1-2}
\cite{li2016modelling} & Identified that learning design has a significant influence on the overall perception of a course. \\
\cmidrule{1-2}
\cite{LAforLDreview} & In a literature review, it was identified that connecting learning design with learning analytics is important for aligning both concepts to gather insights and improve decisions in learning design. \\
\cmidrule{1-2}
\cite{hernandez2019analyticslearningdesign} & Proposed a layered framework that considers analytics layers for learning design. \\
\cmidrule{1-2}
\cite{beyondbias} & Studied the interplay between student satisfaction measurements, workload, and performance. \\
\cmidrule{1-2}
\cite{heffernan2023abusive} & Demonstrated the influence of the gender of the professor on student satisfaction. \\
\cmidrule{1-2}
\cite{professorAgeandGender} & Analyzed the influence of age on student satisfaction. \\
\cmidrule{1-2}
\cite{perceptionsandbiasingfactors} & Identified that cultural and demographic factors potentially influence student satisfaction. \\
\cmidrule{1-2}
\cite{genderandculturalbias} & Demonstrated that gender and cultural biases are present in student satisfaction surveys, finding biases against women and non-native English speakers. \\

\end{longtable}
}
\end{landscape}

\begin{landscape}
\begin{table}[p]
\centering
\caption{Institutional Analytics Framework}
\label{frameworkTable}
\normalsize
\renewcommand{\arraystretch}{1.25}
\setlength{\tabcolsep}{4pt}
\begin{tabular}{L{3.2cm} L{4.8cm} L{3.9cm} L{6.6cm} L{6.7cm}}
\toprule
\rowcolor{gray!15}
\textbf{Factor} & \textbf{Description of the factor} & \textbf{Sub-factor} &
\textbf{Description of the sub-factor} & \textbf{Examples} \\
\midrule

\multirow{5}{=}{\textbf{Analytics Pipeline Factors}} &
\multirow{5}{=}{Process related to the institutional analytics approach applied in a determined problem.} &
Data quality &
Processes related to gathering, preparing, and structuring data for analysis (pre-processing). &
Data collection, feature extraction, sample, data reliability, level of data aggregation. \\
\cmidrule(l){3-5}
 & &
Types of data analysis &
Approaches and assumptions used to interpret and extract insights from data. &
Selection of analysis techniques, visualization, analytical context, assumptions of the data analyses. \\
\midrule

\multirow{7}{=}{\textbf{Institutional Factors}} &
\multirow{7}{=}{Factors related to the institutional level in Higher Education Institutions.} &
Academic centers &
Structural and demographic characteristics of academic departments and centers. &
Academic centers (e.g., Economics, Engineering, Humanities); department gender composition (distribution of professors by gender in each department). \\
\cmidrule(l){3-5}
 & &
Educational framework &
Guidelines for teachers in higher education. &
Teaching methodologies framework. \\
\cmidrule(l){3-5}
 & &
Modality of the university &
Refers to the different ways a university operates. &
Brick-and-mortar university or online university. \\
\midrule

\multirow{6}{=}{\textbf{Course Factors}} &
\multirow{6}{=}{Factors related to the course level in Higher Education Institutions.} &
Course level, class size &
Aspects regarding the course. &
Class size, level of the course. \\
\cmidrule(l){3-5}
 & &
Learning design &
Learning design decisions made in the course. &
Content, assessments, schedule. \\
\cmidrule(l){3-5}
 & &
Students' workload &
Students' satisfaction with workload. &
Students' workload. \\
\cmidrule(l){3-5}
 & &
Student performance &
Student performance. &
Student performance, such as grades. \\
\midrule

\multirow{4}{=}{\textbf{Demographic Factors}} &
\multirow{4}{=}{Factors related to demographic information of students and professors.} &
Gender &
Aspects regarding the influence of gender. &
Gender of the students and gender of the professors. \\
\cmidrule(l){3-5}
 & &
Age &
Aspects regarding the influence of age. &
Age of the professor. \\
\cmidrule(l){3-5}
 & &
Language &
Aspects regarding the influence of language. &
Language of the course. \\
\bottomrule
\end{tabular}
\end{table}
\end{landscape}

\clearpage


\begin{landscape}
\begin{table}[p]
\centering
\caption{System Usability Survey}
\label{tab:survey}
\normalsize
\renewcommand{\arraystretch}{1.25}
\setlength{\tabcolsep}{4pt}
\begin{tabular}{C{1.6cm} L{24.2cm}}
\toprule
\rowcolor{gray!15}
\textbf{No.} & \textbf{Item} \\
\midrule
1  & Did the system help you to get a more responsible interpretation? \\
\cmidrule{1-2}
2  & I think that I would like to use this system frequently \\
\cmidrule{1-2}
3  & I found the system unnecessarily complex \\
\cmidrule{1-2}
4  & I thought the system was easy to use \\
\cmidrule{1-2}
5  & I think that I would need the support of a technical person
   to be able to use this system \\
\cmidrule{1-2}
6  & I found the various functions in this system were well integrated \\
\cmidrule{1-2}
7  & I thought there was too much inconsistency in this system \\
\cmidrule{1-2}
8  & I would imagine that most people would learn to use this system
   very quickly \\
\cmidrule{1-2}
9  & I found the system very difficult to use \\
\cmidrule{1-2}
10 & I felt very confident using the system \\
\cmidrule{1-2}
11 & I needed to learn a lot of things before I could get going
   with this system \\
\bottomrule
\end{tabular}
\end{table}
\end{landscape}


\begin{landscape}
{\normalsize
\renewcommand{\arraystretch}{1.25}
\setlength{\tabcolsep}{4pt}

\RaggedRightParindent=0pt
\RaggedRightRightskip=0pt plus 2em

\newlength{\colA}\setlength{\colA}{0.15\linewidth}
\newlength{\colB}\setlength{\colB}{0.18\linewidth}
\newlength{\colC}\setlength{\colC}{\dimexpr\linewidth-\colA-\colB-4\tabcolsep\relax}

\setlength{\LTleft}{0pt}
\setlength{\LTright}{0pt}
\setlength{\LTcapwidth}{\linewidth}

\begin{longtable}{@{}>{\RaggedRight\arraybackslash}p{\colA}>{\RaggedRight\arraybackslash}p{\colB}>{\RaggedRight\arraybackslash}p{\colC}@{}}
\caption{Coding Scheme for user interactions.}
\label{codingScheme} \\
\toprule
\rowcolor{gray!15}
\textbf{Code} & \textbf{SubCode} & \textbf{Definition and Examples} \\
\midrule
\endfirsthead

\toprule
\rowcolor{gray!15}
\textbf{Code} & \textbf{SubCode} & \textbf{Definition and Examples} \\
\midrule
\endhead

\multicolumn{3}{r}{\small\textit{Continued on next page}} \\
\endfoot

\bottomrule
\endlastfoot

\textbf{Agreement}
  & \textbf{Confirmation}
  & \textbf{A consensus or affirmation regarding a specific statement or piece of information.} User: \textit{``Yes, of course, the score could not be representative if the questionnaire is filled by a few students.''} Assistant: \textit{``Yes, the `Bottom Courses' are those with the lowest student enrollment.''} \\
\cmidrule{2-3}
  & \textbf{Validation}
  & \textbf{Acknowledgment that the provided information is accurate.} Assistant: \textit{``That is an interesting observation about how student and professor gender might relate to satisfaction scores.''; ``Your interpretation effectively explores how the gender of both students and professors might influence satisfaction results across the faculties.''} \\
\midrule

\textbf{Disagreement}
  &
  & \textbf{When there is a disagreement between the user and the assistant regarding a statement or piece of information.} User: \textit{``I don't think the academic unit influences, but the course factors.''; ``there is no information on the course level.''} \\
\midrule

\textbf{Express Doubts}
  &
  & \textbf{When there is ambiguity/doubts about the meaning of a statement or piece of information for either side.} User: \textit{``I don't understand what you mind''; ``The visualization is comparing the shannon index by teaching methodologies, but for me it is not clear the objective of the visualization.''} \\
\midrule

\textbf{Provide Information (PI)}
  & \textbf{Clarification}
  & \textbf{When either side clarifies a statement or piece of information.} Assistant: \textit{``Data Quality examines whether the aggregated scores accurately represent the full range of individual satisfaction responses.''}; User: \textit{``just faculty types.''} \\
\cmidrule{2-3}
  & \textbf{Explanation}
  & \textbf{When there is an explanation from either side about a statement or piece of information.} Assistant: \textit{``The way a methodology is implemented certainly affects the relationship between workload and student performance.'';} User: \textit{``On the other hand for the explorative purposes it is indeed helpful such presentation.''} \\
\cmidrule{2-3}
  & \textbf{Directing Attention}
  & \textbf{When there is a statement or piece of information guiding the user in a different direction of reasoning.} Assistant: \textit{``It is helpful to start by observing the main elements presented in the visualization.''; ``The visualization provides both individual data points (swarm plot) and aggregated summaries (box plot).''} \\
\cmidrule{2-3}
  & \textbf{Reminder}
  & \textbf{When either side reminds the other about a statement or piece of information.} Assistant: \textit{``While instrument testing is crucial, remember that the visualization relies on mean satisfaction scores, which are averages of many individual student responses.''; ``It is important to remember that the Shannon Index calculation relies on the quality of the underlying data regarding resources and activities.''} \\
\cmidrule{2-3}
  & \textbf{Caution}
  & \textbf{When either side expresses caution about a piece of information or statement.} User: \textit{``However these results should be interpreted cautiously as there are outliers include in the analysis and seems like the number of female professors evaluated dont match the male professors evaluated''; ``Acknowledging the potential for bias across different contexts is crucial when assessing Data Quality.''} \\
\cmidrule{2-3}
  & \textbf{Asking for Info}
  & \textbf{When there's a question about a piece of a statement or information.} User: \textit{``Do you mean that the median cannot be representative if, for example, the sample size is small?''; ``How can it be that the average in some cases is lower than the individuals marks?''} \\
\cmidrule{2-3}
  & \textbf{Sharing Insight}
  & \textbf{When there's an observation, thought, or knowledge about a piece of a statement or information.} User: \textit{``Economics faculty: female students seem to be more satisfied with the teaching they received from female teachers, and display a higher variability in the teaching received from male teachers.''}; Assistant: \textit{``Different teaching methodologies often require varying levels of student effort and time commitment.''} \\
\cmidrule{2-3}
  & \textbf{Statement}
  & \textbf{When there's a piece of information.} Assistant: \textit{``Your interpretation relies on the assumption that the data accurately reflects the course quality.''}; User: \textit{``Having both results (individual scores and the group average) is very helpful''} \\
\cmidrule{2-3}
  & \textbf{Summary}
  & \textbf{When there's a summary of the conversation.} Assistant: \textit{``Here is a summary for Performance and Workload by Active Teaching Methodology. This analysis of teaching methodologies, performance, and workload requires careful consideration of several factors for a responsible interpretation.''; ``Here is a summary for Diversity of LMS configuration by Teaching Methodology. This visualization compares the variety of LMS resources and activities, measured by the Shannon Index, across Traditional Teaching and Active Learning methodologies.''} \\
\midrule

\textbf{Request Information (RI)}
  & \textbf{Factors}
  & \textbf{When there's a clear question or information regarding the framework's factors.} Assistant: \textit{``Now that we have considered the data itself, how might the specific context of Demographic Factors --- Gender influence your interpretation of the satisfaction differences?''; ``How could the specific Learning Design of these methodologies influence whether students feel the workload is justified by the results?''} \\
\cmidrule{2-3}
  & \textbf{Framework}
  & \textbf{When there's a question about the framework.} Assistant: \textit{``Considering the framework, how might the interaction between the student's gender and the professor's gender affect the resulting satisfaction scores?''} \\
\cmidrule{2-3}
  & \textbf{Reflection}
  & \textbf{When a question guides either side through a line of reasoning.} Assistant: \textit{``How might this specific workload difference influence the performance gap you identified for FLIP?''}; User: \textit{``So that it can be a bias?''} \\
\cmidrule{2-3}
  & \textbf{Interpretation}
  & \textbf{When a question directs either side along an interpretation.} Assistant: \textit{``How do you think the Learning Design, such as the type of tasks or LMS usage, might influence these outcomes, and how might this relate to the Student's workload?''}; User: \textit{``It can be that one type of methodology has been more implemented in one department and less in other?''} \\
\cmidrule{2-3}
  & \textbf{Directing Attention}
  & \textbf{When a question leads either side to a new line of reasoning.} Assistant: \textit{``Now, examine the workload data for FLIP: is there a notable difference in the reported workload between the top and bottom students?''; ``Now, let's shift our focus to the Analytics Pipeline Factors: how might the Data Quality influence the reliability of these performance and workload measurements?''} \\
\cmidrule{2-3}
  & \textbf{Visualization}
  & \textbf{When there's a direct question about the visualization.} Assistant: \textit{``How does the use of a weighted average specifically impact how you interpret the performance differences between the methodologies shown in the visualization?''; ``What does the height or spread of the boxes suggest to you about the variety of LMS configurations within each teaching methodology?''} \\
\cmidrule{2-3}
  & \textbf{Asking for Clarification}
  & \textbf{When there's a question asking for clarification.} User: \textit{``Well, if data have been standardized, the mean should be calculated once it has been done, isn't it?''; ``Lectures has more index than others in the top courses, why is it?''} \\
\midrule

\textbf{Reflection}
  &
  & \textbf{When there is reasoning based on the interpretation of the visualization.} User: \textit{``But at the end of the course, I think good students evaluate methodologies that allowed them to increase their knowledge.''; ``In my opinion, project-based learning methodologies are hands-on learning situations and experiences, usually made in group, so team-work is a big factor influencing the outcome.''} \\
\end{longtable}
}
\end{landscape}


\begin{table}[htbp]
\centering
\caption{Number of sub-factors addressed without and with the use of the FACTRIA-aware chatbot.}
\label{tab:factria_results}
\begin{tabular}{clcc}
\toprule
\rowcolor{gray!15}
\textbf{Participant} & \textbf{Assigned case (max number of sub-factors)} & \textbf{Without} & \textbf{With} \\
\midrule
\multirow{2}{*}{P1}  & CASE 3 (5 sub-factors) & 1 & 3 \\
                     & CASE 4 (4 sub-factors) & 0 & 4 \\
\midrule
\multirow{2}{*}{P2}  & CASE 1 (4 sub-factors) & 2 & 4 \\
                     & CASE 4 (4 sub-factors) & 1 & 4 \\
\midrule
\multirow{2}{*}{P3}  & CASE 1 (4 sub-factors) & 1 & 2 \\
                     & CASE 2 (5 sub-factors) & 0 & 3 \\
\midrule
\multirow{2}{*}{P4}  & CASE 1 (4 sub-factors) & 2 & 4 \\
                     & CASE 4 (4 sub-factors) & 0 & 4 \\
\midrule
\multirow{2}{*}{P5}  & CASE 2 (5 sub-factors) & 2 & 4 \\
                     & CASE 3 (5 sub-factors) & 1 & 1 \\
\midrule
\multirow{2}{*}{P6}  & CASE 2 (5 sub-factors) & 3 & 5 \\
                     & CASE 3 (5 sub-factors) & 1 & 3 \\
\midrule
\multirow{2}{*}{P7}  & CASE 1 (4 sub-factors) & 1 & 3 \\
                     & CASE 2 (5 sub-factors) & 2 & 3 \\
\midrule
\multirow{2}{*}{P8}  & CASE 3 (5 sub-factors) & 1 & 3 \\
                     & CASE 4 (4 sub-factors) & 1 & 4 \\
\midrule
\multirow{2}{*}{P9}  & CASE 1 (4 sub-factors) & 2 & 4 \\
                     & CASE 4 (4 sub-factors) & 2 & 4 \\
\midrule
\multirow{2}{*}{P10} & CASE 1 (4 sub-factors) & 2 & 3 \\
                     & CASE 2 (5 sub-factors) & 2 & 4 \\
\midrule
\multirow{2}{*}{P11} & CASE 1 (4 sub-factors) & 2 & 4 \\
                     & CASE 3 (5 sub-factors) & 2 & 3 \\
\bottomrule
\end{tabular}
\end{table}

\clearpage
\restoregeometry   

\section{FACTRIA-Aware Chatbot: Technical Details}
\label{techDetails}
  
Each API request incorporates three structured elements: (1) a \texttt{<CONTEXT>} block with the visualization's title and descriptive text explaining what the chart represents, the underlying data, the variables displayed, and relevant methodological notes (e.g., standardization procedures, data aggregation methods); (2) a \texttt{<FACTOR\_STATE>} block specifying the current factor from FACTRIA; and (3) a \texttt{<FRAMEWORK>} block containing the FACTRIA entries for that case. These XML-like tags mark the injected content as the authoritative source for the response. The system assembles context through direct injection. For each case, two researchers identified the FACTRIA entries relevant to the visualization in advance, and the system supplies those entries, the case description, and the tabular values underlying the chart verbatim in every request. This fixed, auditable evidence set therefore bounds generation instead of relying on a query-time search.

Three mechanisms limit unsupported statements: the injected blocks restrict response content; the model may reference numerical values only when they appear in the injected data; and near-deterministic decoding reduces variability. We did not measure adherence quantitatively, and we discuss the residual risk among the limitations.

The interaction advances through a deterministic state machine, and an auxiliary classification call selects each transition. The call classifies the first user message as INITIAL\_INTERPRETATION, GREETING\_OR\_SMALLTALK, or READY\_BUT\_VAGUE. It then classifies each subsequent message, relative to the factor under discussion, as SUBSTANTIVE (advance to the next factor), EXPLORING (re-prompt on the same factor), CLARIFICATION (hand over to the clarification agent), or END (proceed to the summary). The clarification agent returns a two- to three-sentence definition of at most 80 words. Five prompted components therefore make up the system: an opening-intent classifier, a turn classifier, a clarification agent, the primary agent, and the summarization agent.

The primary agent follows an instruction set designed to elicit user reflection. We instruct the primary agent to respond in at most two short sentences (50 words) and avoid technical jargon. The agent never provides direct interpretations of the visualization. We also instruct it to redirect users who overlook or avoid a factor.

After the dialogue ends, the summarization agent generates a 150-word summary for practitioners. It receives the full conversation and the relevant FACTRIA factors, producing a critical yet constructive synthesis that highlights the factors discussed and gently corrects misunderstandings. The tone remains professional and accessible, framing limitations through expressions such as ``a responsible interpretation could help us understand\ldots''. A JSON schema constrains the output and records, for each factor discussed, a coverage label (thorough, partial, or surface) and the insight the user reached, together with the factors not reached and a short narrative.

All components use Google's Gemini API (\texttt{gemini-2.0-flash}) with conservative settings (temperature = 0.2, top-p = 0.9) to promote coherence and reduce variability. We cap the conversational agent at 200 tokens to maintain brevity and allocate up to 2,048 tokens to the summarization agent for adequate synthesis. The interface operates in English, Spanish, and Catalan, and we instruct all agents to reply in the participant's language. The prompts and the FACTRIA entries used by the system are available at \url{https://doi.org/10.5281/zenodo.23082579}.

\clearpage

\section{Transition Network Analysis}
\label{AppendixTNA}
This section presents case-specific transition networks. Each network includes transitions observed at least twice with a probability greater than 0.2. 

\subsection{Case 1}

\begin{figure}[hbt!]
    \centering
    \includegraphics[width=\columnwidth]{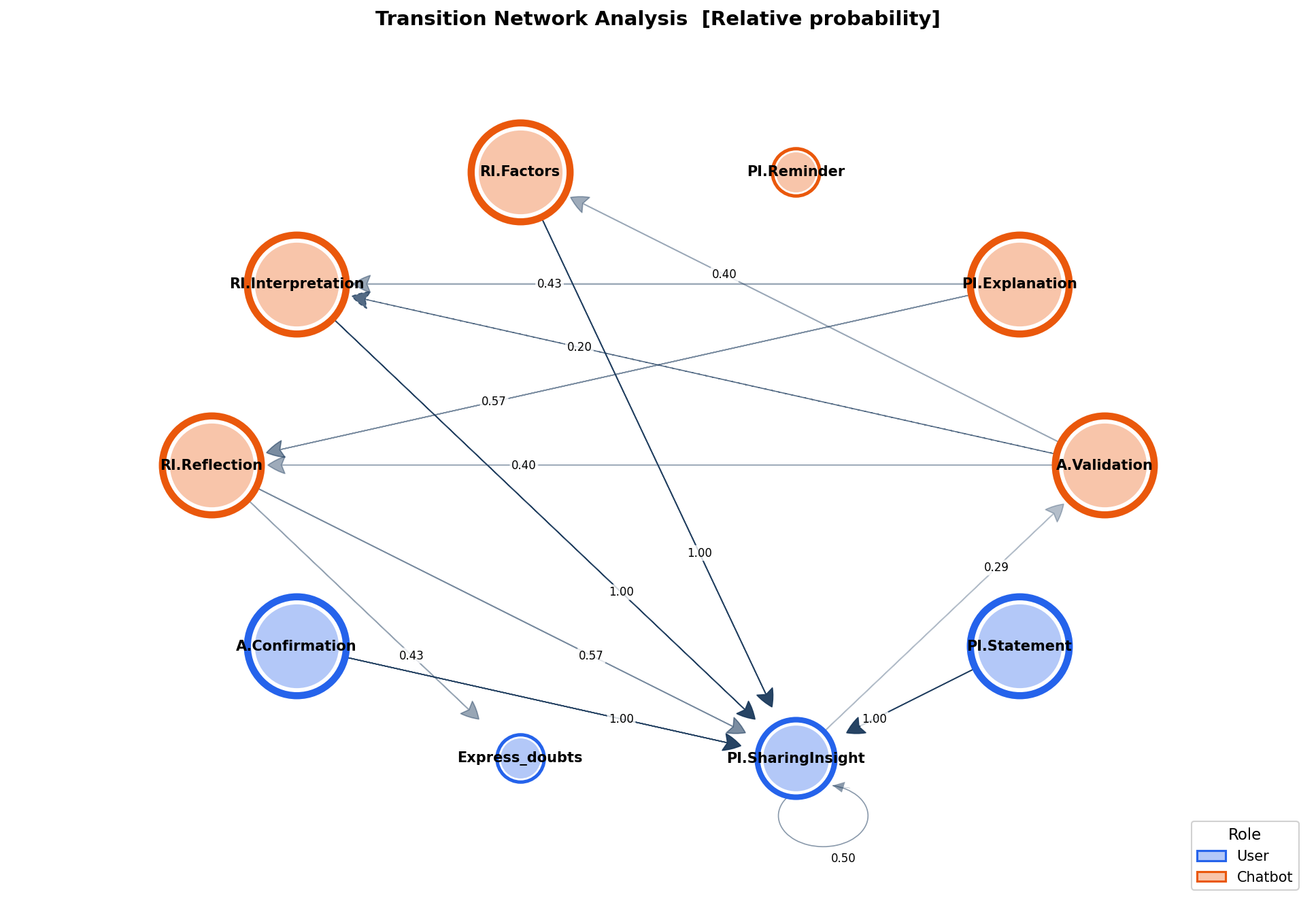}
    \caption{Transition Network Analysis Case 1.}
    \label{fig:TNACase1}
\end{figure}

Figure \ref{fig:TNACase1} shows that the assistant guides the user towards responsible interpretation through prompts coded as \textit{RI.Reflection}, \textit{RI.Interpretation}, and \textit{RI.Factors}; some corresponding transitions have a probability of $P = 1$. The user subsequently shares insights (\textit{PI.SharingInsight}) or expresses agreement (\textit{A.Confirmation}).

\clearpage

\subsection{Case 2}

\begin{figure}[hbt!]
    \centering
    \includegraphics[width=\columnwidth]{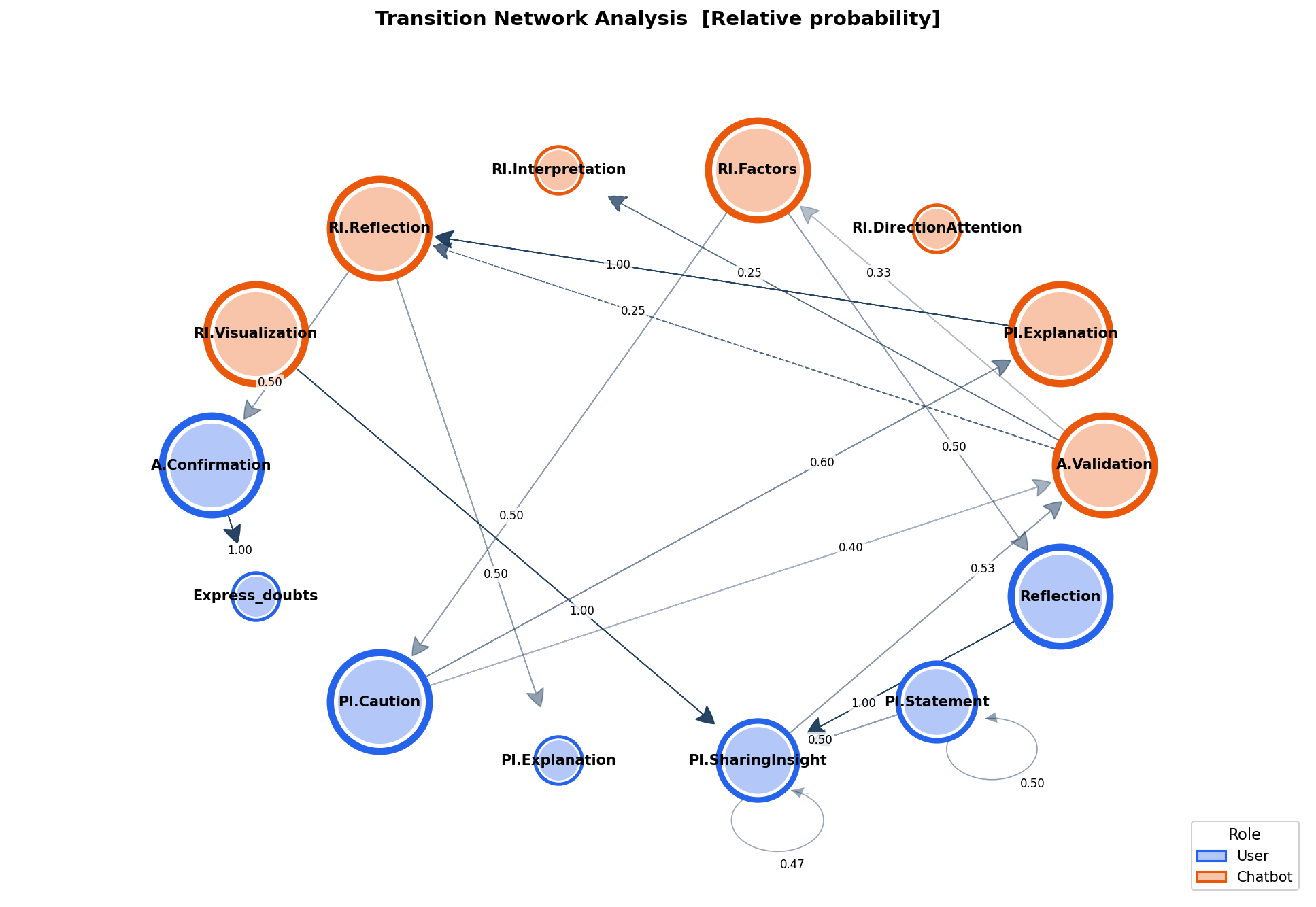}
    \caption{Transition Network Analysis Case 2.}
    \label{fig:TNACase2}
\end{figure}

Figure \ref{fig:TNACase2} shows a more diverse set of user–assistant interactions. The user contributes actions coded as \textit{PI.Caution}, \textit{Reflection}, and \textit{PI.SharingInsight}, while the assistant contributes prompts coded as \textit{RI.Reflection}, \textit{RI.Visualization}, and \textit{RI.Factors}. The higher-probability transitions indicate that \textit{PI.Explanation} follows assistant \textit{RI.Reflection}, while user \textit{Reflection} follows \textit{PI.SharingInsight}.

\clearpage

\subsection{Case 3}

\begin{figure}[hbt!]
    \centering
    \includegraphics[width=\columnwidth]{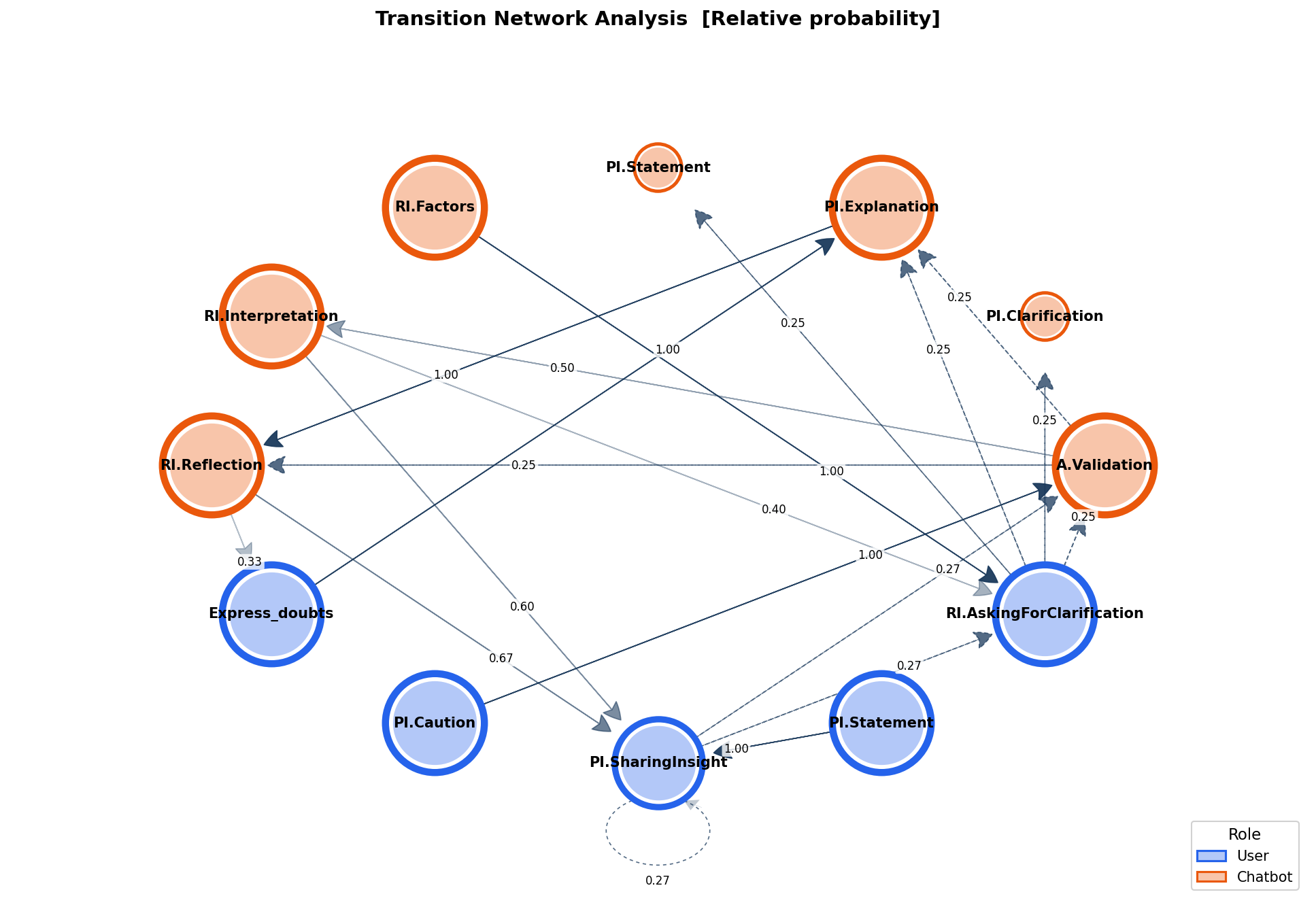}
    \caption{Transition Network Analysis Case 3.}
    \label{fig:TNACase3}
\end{figure}

Figure \ref{fig:TNACase3} shows diverse interactions between the user and assistant, consistent with guidance rather than answer provision. The user contributes actions coded as \textit{PI.SharingInsight}, \textit{RI.AskForClarification}, \textit{ExpressDoubts}, and \textit{PI.Caution}. The assistant asks reflective questions coded as \textit{RI.Factors}, \textit{RI.Interpretation}, and \textit{RI.Reflection}, and explains or validates user contributions through \textit{PI.Explanation} and \textit{A.Validation}.

\clearpage

\subsection{Case 4}

\begin{figure}[hbt!]
    \centering
    \includegraphics[width=\columnwidth]{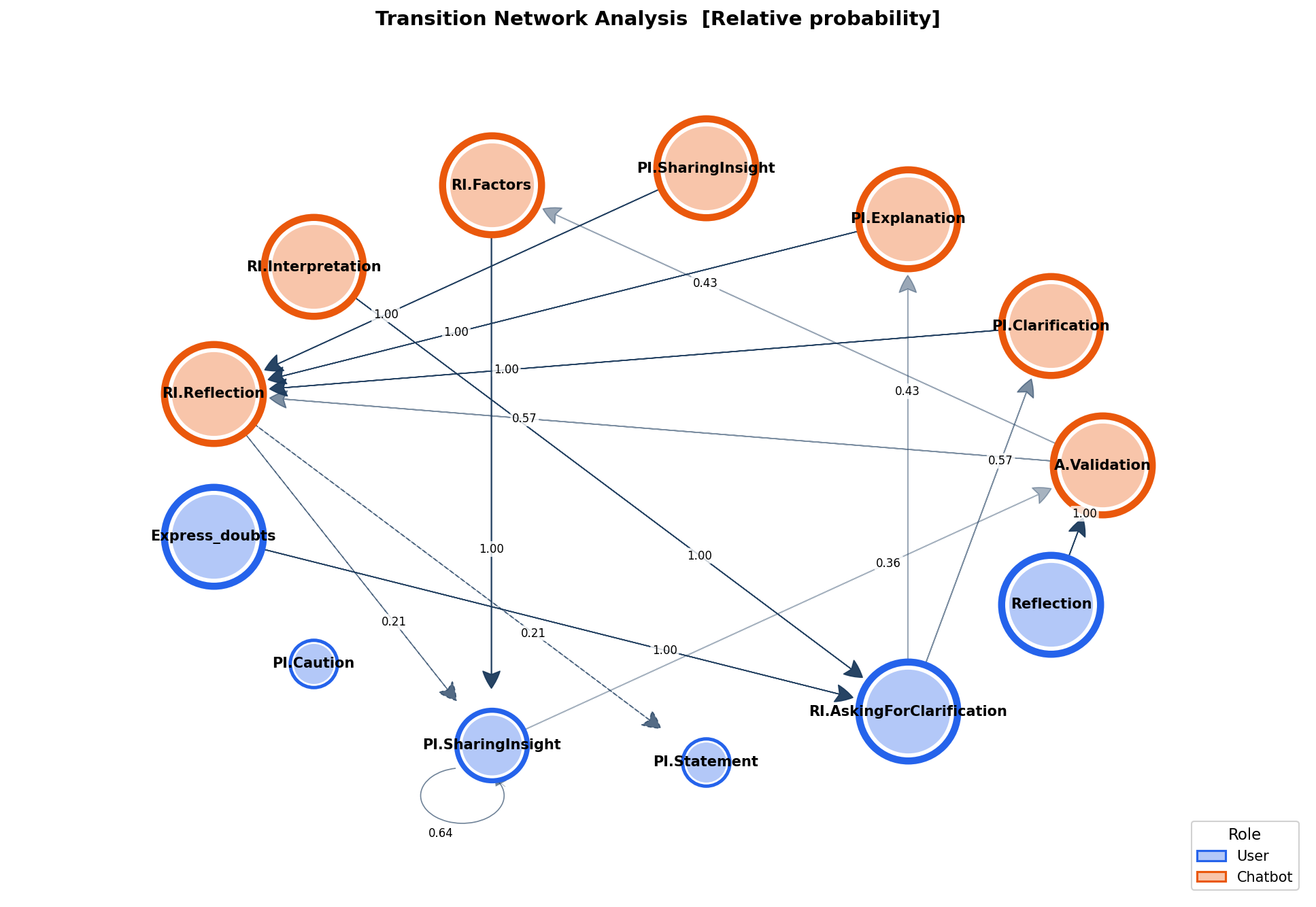}
    \caption{Transition Network Analysis Case 4.}
    \label{fig:TNACase4}
\end{figure}

Figure \ref{fig:TNACase4} shows that prompts coded as \textit{RI.Factors}, \textit{RI.Interpretation}, and \textit{RI.Reflection} precede the user's \textit{PI.SharingInsight}. Following \textit{RI.Reflection}, the assistant also contributes \textit{PI.Explanation}, \textit{PI.Clarification}, or \textit{PI.SharingInsight}. When the user expresses doubt or requests clarification (\textit{ExpressDoubts}; \textit{RI.AskForClarification}), the assistant responds with \textit{PI.Clarification} followed by another \textit{RI.Reflection} prompt.  

\clearpage

\end{document}